# Small Near-Earth Asteroids as a Source of Meteorites

**Jiří Borovička and Pavel Spurný**
*Astronomical Institute of the Czech Academy of Sciences*

**Peter Brown**
*University of Western Ontario*

___

Small asteroids intersecting Earth's orbit can deliver extraterrestrial rocks to the Earth, called meteorites. This process is accompanied by a luminous phenomena in the atmosphere, called bolides or fireballs. Observations of bolides provide pre-atmospheric orbits of meteorites, physical and chemical properties of small asteroids, and the flux (i.e. frequency of impacts) of bodies at the Earth in the centimeter to decameter size range. In this chapter we explain the processes occurring during the penetration of cosmic bodies through the atmosphere and review the methods of bolide observations. We compile available data on the fireballs associated with 22 instrumentally observed meteorite falls. Among them are the heterogeneous falls Almahata Sitta (2008 $TC_3$) and Benešov, which revolutionized our view on the structure and composition of small asteroids, the Příbram-Neuschwanstein orbital pair, carbonaceous chondrite meteorites with orbits on the asteroid-comet boundary, and the Chelyabinsk fall, which produced a damaging blast wave. While most meteoroids disrupt into fragments during atmospheric flight, the Carancas meteoroid remained nearly intact and caused a crater-forming explosion on the ground.

## 1. INTRODUCTION

Well before the first asteroid was discovered, people unknowingly had asteroid samples in their hands. Could stones fall from the sky? For many centuries, the official answer was: NO. Only at the end of the 18$^{th}$ century and the beginning of the 19$^{th}$ century did the evidence that rocks did fall from the sky become so overwhelming that this fact was accepted by the scientific community. It was the German scientist Ernst Chladni who first recognized their extraterrestrial origin in 1794. Nevertheless, it took some time before his idea became accepted (see e.g. *Lauretta and McSween,* 2006).

The stones that fell from the sky are distinct from all terrestrial rocks. They are called *meteorites*. According to their mineralogical composition they are classified into many different types. The two basic classes are stony and iron meteorites, although a mixture of these chemistries (stony-iron meteorites) exists as well. Iron meteorites are composed of metallic iron with an admixture of nickel and some traces of other elements. Stony meteorites are mostly composed of silicates such as olivine and pyroxene, with some metallic iron also present in many cases. The elements and most of minerals found in meteorites are known from terrestrial rocks but their ratios in meteorites are different. The most common type of stony meteorites, called chondrites, contains mm-sized spherical structures called chondrules. Such structures are not present in any terrestrial rock. More detailed classification of meteorites is given in Binzel et al. (this volume). The study of meteorites is a well developed scientific field and rich literature exists (e.g. *Dodd,* 1981*; Papike,* 1998*; Hutchison,* 2004*; Lauretta and McSween,* 2006).

According to the circumstances of their recovery, meteorites can be divided into *finds* and *falls*. Finds are meteorites found on ground (by chance or by dedicated searches) but the date of their fall is unknown. Falls are meteorites, whose fall was witnessed – either the meteorite was seen (or heard) to hit the ground or the fireball caused by the ablation of the parent meteoroid in the atmosphere was observed and the meteorites were found later. *Meteoroid* is a term for the original body, which dropped the meteorite(s). In fact, there is no strictly defined boundary between large meteoroids and small asteroids. For the purpose of this chapter, we will call meteoroids all bodies smaller than 10 meters and larger than 50 micrometers. Bodies smaller than 50 micrometers are called *dust particles*. A dust particle, when entering the atmosphere, may not give rise to the phenomenon called a meteor, depending on its initial speed. *Meteor* is the radiation, and associated phenomenon (heat, shock, ionization), caused by the entry of a meteoroid in the atmosphere. Dust particles are sufficiently small that they may decelerate before they start to evaporate, so they do not produce meteors. They gradually sediment through the atmosphere and finally reach the ground.

The vast majority of meteoroids are destroyed in the atmosphere and do not produce any macroscopic meteorites. Parts of them may not be evaporated completely and can reach the ground as dust or *micrometeorites* (objects smaller than ~one millimeter). To produce a meteorite, the meteoroid must be large enough, mechanically strong enough, and have relatively low entry velocity. The initial velocities of meteoroids at entry to the terrestrial atmosphere in bound heliocentric orbits are between 11 and 71 km/s. The chance of surviving the passage through the atmosphere strongly decreases with increasing velocity. As a rough empirical rule (to which are many exceptions), a meteoroid of stony composition with low entry velocity (~ 20 km/s) can produce meteorites of total mass of 5 – 10% of the initial (pre-atmospheric) mass of the meteoroid (*Halliday et al.*, 1989a; *Popova et al.*, 2011)

A meteor brighter than stellar magnitude − 4 (i.e. brighter than planet Venus) is called a *fireball* or *bolide*. The term *superbolide* is used for meteors brighter than magnitude −17 (Ceplecha et al., 1998). Meteorite falls are always in the bolide or superbolide category. Nevertheless, it is not true that the brighter the bolide the larger the meteorite. There are fluffy meteoroids of cometary origin, which produce bright bolides but are completely destroyed high in the atmosphere (e.g. *Borovička and Spurný* 1996, *Madiedo et al.* 2014a).

Meteorites, as extraterrestrial samples, have much scientific value. However, to fully exploit the information they contain, it is necessary to know where they come from. Early researchers thought meteorites may come from interstellar space (see e.g. the discussions of *Öpik*, 1950, and *Fessenkow et al.* 1954). It was the Příbram meteorite, which fell in the former Czechoslovakia in 1959 and whose trajectory, impact position, and orbit was determined rigorously for the first time from bolide photographs, which placed meteorites in the context of the Solar System (*Ceplecha,* 1961). Since bolides are rare events and occur unexpectedly, it is necessary to monitor systematically large territories to obtain observational data. For that reason, dedicated arrays of cameras, called fireball networks, were put in operation in several countries. One of their goals was to assist with meteorite recovery and, at the same time, provide orbits of the meteorites (*Ceplecha,* 1986). In more modern times, when various types of still and video cameras became widespread among the public, casual instrumental records of bolides have become common. These records, after careful calibration (*Borovička,* 2014) and often laborious computation, also provided some meteorite orbits, though with lower precision than dedicated bolide cameras. As of 2014, there are 22 meteorites with reliably (i.e. instrumentally determined) known pre-impact orbit (see Section 5). The orbits confirm that most meteorites are fragments of asteroids, although a cometary origin of a small fraction of meteorites cannot be fully excluded. In addition, it was found that some meteorites, according to their composition, must originate from the Moon and some others from the Mars, though no orbits for lunar or Martian meteorite falls are yet documented.

It is estimated that about 4500 meteorite falls dropping more than 1 kg of total meteorite mass occur annually on the Earth (*Halliday et al.,* 1989b). Only in one case has the meteorite fall been predicted in advance: a small interplanetary object designed as 2008 TC$_3$ was discovered on October 7, 2008 when approaching the Earth and predicted to enter the atmosphere the next day over Sudan. Its remnants were later recovered as the Almahata Sitta meteorites (*Jenniskens et al.,* 2009). On January 1, 2014, another object, 2014 AA, was discovered on an Earth-approaching orbit but the impact point was poorly constrained. There are indications that the impact occurred into the Atlantic Ocean (*Beatty,* 2014).

The observations of bolides, with or without an associated meteorite fall, provide information about the orbital distribution, physical properties, and (when accompanied with spectral observation) composition of fragments of asteroids (and comets) intersecting the Earth's orbit. They are therefore a complementary tool to the astronomical and in situ studies of asteroids. They also provide direct information on the effects of impacts on the Earth. Large bodies generate a strong blast wave in the atmosphere, which can have damaging effects on the surface, as was the case of Tunguska event over Siberia in 1908 (e.g. *Vasilyev,* 1998) and Chelyabinsk in 2013 (*Brown et al.,* 2013a; *Popova et al.,* 2013). Even small meteoroids can occasionally cause an impact crater, as was the case for the Carancas impact in 2007 (e.g. *Tancredi et al.,* 2009).

This chapter provides an overview of what is known about bolides producing meteorite falls. Special attention is devoted to instrumentally observed meteorite falls. These are the events with the most complete information available (analysis of both the bolide and the corresponding meteorite). We note that these phenomena are not restricted to the Earth. Meteorites (*Fairén et al.,* 2011), fresh impact craters and crater strewn fields (*Daubar et al.,* 2013) were discovered on the surface of Mars and superbolides have now been observed in the atmosphere of Jupiter (*Hueso et al.,* 2013). On the atmosphere-less Moon, flashes caused by the direct impact of meteoroids on the surface can be observed (*Suggs et al.,* 2014).

## 2. METEOROIDS IN THE ATMOSPHERE

In this section we describe in more detail the penetration of large (> 10 cm) meteoroids thorough the atmosphere. The understanding of these processes is important for the derivation of meteoroid properties from bolide observations, for assessment of the hazard connected with meteoroid/asteroid entry and for improving the efficiency of meteorite recoveries.

In the upper layers of the atmosphere, at heights > 130 km, individual atoms and molecules directly impact the meteoroid surface. This leads to gradual heating of the meteoroid surface. Under some circumstances, the collision with atmospheric molecules can lead to the release of a meteoric atom from the surface. This process, called *sputtering*, is a form of slow mass loss at low temperatures (*Rogers et al.,* 2005). It is efficient only at meteoroid velocities larger than 30 km/s (*Popova et al.,* 2007). Intense mass loss starts only when the meteoroid surface is heated to its melting temperature of about 2000 K (*Ceplecha et al.,* 1998). The molten surface layer partly evaporates and is partly lost in the form of liquid droplets, which then continue to evaporate in the hot plasma surrounding the meteoroid in a process called *thermal*

*ablation*. The hot envelope of heated air and meteoric vapors around the meteoroid is the main source of bolide radiation. The plasma temperature and composition can be studied by the methods of meteor spectroscopy. The typical temperatures are 4000 – 5000 K (e.g. *Borovička,* 2005). At these temperatures, the main contributors to the bolide radiation in visible light are meteoric metals, in particular Fe, Mg, Na, Ca, Cr, Mn. They radiate in the form of atomic emission lines. The atmospheric species (N, O) and some meteoric species (notably Si, S, C) have no bright lines in the visible range. As a result, their abundance is difficult to determine. Moreover, the composition of meteoric vapors usually does not fully reflect the composition of the meteoroid as some refractory elements, notably Al, Ca, Ti, are not evaporated completely during thermal ablation particularly at heights above 30 km (*Borovička and Spurný,* 1996; *Borovička,* 2005). Nevertheless, bolide spectroscopy can be used to distinguish the main types of meteoritic material (chondrites, achondrites, irons) in the cases when no meteorite is recovered. There is also a component of higher temperature (~10,000 K) observed in bolide spectra (*Borovička,* 1994a). The strength of this second component (represented mainly by the lines of $Ca^+$, $Mg^+$, and $Si^+$) increases rapidly with bolide velocity and is not seen in slow bolides. The high temperature region is probably formed in front of the body, where the interaction of the ablated material and the air flux is the strongest.

In the denser atmospheric layers, atmospheric molecules do not impact directly the meteoroid surface. The *free molecular regime* changes into the *continuous flow* (*Popova,* 2004*)*, which can be treated within the framework of gas hydrodynamics. A shock wave forms at the boundary between the envelope protecting the meteoroid and the incoming flow of the atmosphere. According to models, the temperature of the shock heated air can reach tens of thousands degrees (*Artemieva and Shuvalov,* 2001).

The interaction with the atmosphere leads to loss of mass, deceleration, and in many cases to the fragmentation of the meteoroid. The classical meteor equations describe the deceleration and mass loss (ablation) of a single (non-fragmenting) meteoroid (*Bronshten,* 1983; *Ceplecha et al*., 1998). In this treatment, the meteoroid with mass *m*, cross-section *S,* moving with a velocity *v,* encounters atmosphere of mass $S\rho v dt$, where ρ is the density of atmosphere and dt is the considered time interval. The momentum of the encountered atmosphere is $S\rho v^2 dt$ and the kinetic energy is $½ S\rho v^3 dt$. From the conservation of momentum, we have the *drag equation* describing deceleration of the meteoroid:

$$m\, dv/dt = -\Gamma S\rho v^2,$$

where Γ (sometimes written as $C_D/2$) is the *drag coefficient*. The energy is consumed in mass loss according to the *ablation equation*:

$$dm/dt = -\Lambda S\rho v^3/2Q,$$

where Λ is the *heat transfer coefficient* and *Q* is the energy necessary to ablate a unit mass of the meteoroid. These equations are usually rewritten by introducing the *shape factor* $A = Sm^{-2/3}\delta^{2/3}$, where δ is the bulk density of the meteoroid, and the *ablation coefficient* is $\sigma = \Lambda/2Q\Gamma$, the latter of which can be directly inferred from observations. More details and the analytical integrals of the above equations (assuming constant coefficients) can be found in *Ceplecha et al.* (1998). Theoretical models of meteoroid ablation were discussed e.g. by *Baldwin and Schaeffer* (1971), *Biberman* (1980), *Svetsov et al.* (1995), *Golub et al.* (1996), and *Nemtchinov et al.* (1997).

The single body approach is rarely applicable for the entirety of the bolide's flight. Meteoroid fragmentation in the atmosphere is a complex process, which is not possible to predict exactly but is ubiquitous. It depends on the structural properties of each individual meteoroid and occurs in various forms, i.e. chipping off a small part of the body, disruption into two or more fragments of similar sizes, catastrophic disruption into large numbers of small fragments etc. Fragmentation is often a multi-stage process, whereby fragments arising from early fragmentation episodes disrupt again later. If this process has some regularity, it is called *progressive fragmentation*. The term *quasi-continuous fragmentation* refers to a process when small fragments (dust) are released from the main body almost continuously. This is the dominant fragmentation mode for a kind of weak mm-sized population of meteoroids of cometary origin (*Borovička et al.,* 2007).

For a given bolide velocity, the structurally weaker is the meteoroid, the higher in altitude the fragmentation starts. Some types of fragmentation can be induced thermally; nevertheless, in most cases (especially when large fragments are involved or when catastrophic disruption occurs) the breakup is likely due to aerodynamic loading. In this process, the dynamic pressure acting at the front surface of the meteoroid is $p = \Gamma \rho v^2$, while the pressure at the rear side is zero, and the difference in the pressures causes structural failure of the meteoroid. Fragmentation occurs when $p$ exceeds the strength of the meteoroid (see *Holsapple, 2009*, for various definitions of strength). In estimating the dynamic pressure at breakup, the factor $\Gamma$, of the order of unity, is often neglected and the meteoroid strength is estimated as $\rho v^2$. The velocity, $v$, is easily measurable from meteor data, but the determination of the fragmentation height (and thus the corresponding atmospheric density, $\rho$) is often difficult. This can be done by a number of methods (geometric, photometric, dynamic, acoustic), depending on the type of available data and the type of fragmentation (*Ceplecha et al.,* 1993, *Trigo-Rodríguez and Llorca*, 2006, *Popova et al.* 2011)

The radiation of the bolide is assumed to be proportional to the instantaneous loss of kinetic energy, as expressed by the luminosity equation:

$I = - \tau (½ v^2 \, dm/dt + mv \, dv/dt)$,

where I is the radiative output and $\tau$ is the *luminous efficiency* (which may depend on meteoroid velocity, mass, composition, and height in the atmosphere). For cases including fragmentation, the total output is the sum of contributions of all fragments. When a large number of fragments is released, a sudden increase in brightness, called a *flare*, occurs due to the increase of the meteoroid total cross-section.

Thermal ablation stops when the meteoroid velocity decreases below about 3 km/s. The end of the bolide therefore occurs when either all mass has been ablated and no macroscopic fragments remain or the velocity of all fragments decreases below this ablation limit. In the latter case, the fragments continue to fall during a period called *dark flight*. Their surface gradually cools and no light is emitted (except perhaps a faint infrared glow at the beginning). Typically, the velocity drops below the ablation limit at a height somewhere between 10 – 30 km, depending mainly on the initial meteoroid mass and its fragmentation in the atmosphere. The deceleration continues during dark flight. The fragments follow a ballistic trajectory which turns into a nearly vertical fall, influenced by atmospheric winds (*Ceplecha,* 1987). When they reach the ground, the fragments are called *meteorites*. Typical impact speeds are in the range 10 – 100 m/s for meteorites of 0.1 – 100 kg. Fresh meteorites are characterized by a dark fusion crust representing a resolidified layer of molten material acquired during the latter stages of ablation (*Genge and Grady,* 1999). Sometimes a part of the fusion crust is missing, suggesting that fragmentation continues to occur during the dark flight.

In rare cases of large, strong meteoroids, the deceleration may not be sufficient to convert the flight into free fall. If the body hits the ground with a velocity larger than about 0.5 – 1 km/s, it generates a shock wave in the ground resulting in crater formation. An *impact crater* much larger than the impactor is then formed (*Holsapple,* 1993).

The radiation is not the only demonstration of meteoroid flight through the atmosphere. The supersonic flight generates a cylindrical *blast wave*, which can be heard on the ground and, if the amplitude and the seismic characteristics of the ground are appropriate, seismic waves may be excited (*Edwards et al.,* 2008). Meteoroid fragmentation events can generate spherical blast waves. These waves travel in the atmosphere with the speed of sound (~ 300 m/s), so they reach the ground tens of seconds to minutes after the bolide, depending on the range of the bolide. Non-audible sound waves of low frequency, termed *infrasound*, attenuate particularly slowly in the atmosphere and can be detected with special detectors over large distances, in some cases for tens of thousands kilometers (*Ens et al.,* 2012).

Eyewitnesses of bolides sometimes report another type of sound (variously described as a hissing, popping or crackling) heard simultaneously with the bolide. The origin of these *electrophonic sounds* is not well understood but they are believed to be transmitted as VLF/ELF electromagnetic waves and converted into audible sound by the vibration of objects in the vicinity of the observer (*Keay,* 1992). Recently, radio emission at frequencies 20 – 40 MHz was reported from fireball trails (*Obenberger et al.,* 2014).

Bolides, but also quite faint meteors, produce *ionization trails* in the atmosphere. The lifetimes of the trails vary from a fraction of second to many minutes. They reflect electromagnetic radiation, which is used for the detection of meteors by radars (*Ceplecha et al.,* 1998, *Jones et al.,* 2005). High-power, large-aperture (HPLA) radars are able to detect the plasma that forms in the vicinity of tiny meteoroids (*Close et al.,* 2007). For larger bodies, this so called *head echo* can be detected also by normal meteor radars (*Brown et al.,* 2011).

## 3. FIREBALL OBSERVATIONS

In this section we provide a brief overview of bolide observations and the basic methods of data analysis.

### 3.1 Observation methods

Bolides and more so superbolides are conspicuous phenomena on the sky (Fig. 1) and often draw the attention of the public. They can also be detected by various instruments – optical cameras, photoelectric sensors, radars, acoustic and seismic detectors or satellite-based sensors. Optical imaging cameras provide the most straightforward data about the bolide trajectory, velocity, and luminosity. The bolide must be imaged from at least two widely separated (optimally about 100 km apart) sites to reconstruct the trajectory. Inspired by the success with the Příbram bolide, whole networks of cameras have been established to capture similar events and to characterize the population of large meteoroids. The network, which started in Czechoslovakia in 1963 (*Ceplecha and Rajchl,* 1965) and was joined by Germany in 1968 (*Ceplecha et al.,* 1973, *Oberst et al.,* 1998), formed the European Fireball Network, which remains in operation today. At the beginning, the network used low resolution all-sky mirror cameras recording on photographic film. The Czech part has been modernized in several stages following the advancement of technology: mirrors have been replaced by fish-eye lenses providing higher resolution after 1975 (*Ceplecha,* 1986), manually operated cameras were replaced by Autonomous Fireball Observatories, AFO beginning in 2001 (*Spurný et al.,* 2007) and film versions are currently being replaced by digital versions (the Digital AFO or DAFO). The Australian Desert Fireball Network, which has operated since 2005 (*Bland et al.,* 2012), also uses AFO. The Prairie Network operated in the USA from 1963 – 1975 and used batteries of high resolution photographic cameras (*McCrosky and Boeschenstein,* 1965). This was also the case for the MORP (Meteorite Observation and Recovery Project) in Canada active between 1971 – 1985 *(Halliday et al.,* 1978, 1996). The three early networks were compared by *Halliday* (1973). The Tajikistan Fireball Network operated in 2009 – 2012 (two stations have been in operation since 2006 until now) and used manually operated fish-eye film cameras together with digital cameras (*Kokhirova and Borovička,* 2009). There are also networks based on video cameras, either all-sky versions used in the Southern Ontario Meteor Network, operated in Canada since 2004 (*Brown et al.,* 2010), in the NASA fireball network operated in the USA since 2008 (*Cooke and Moser,* 2012), and in the Slovak Video Meteor Network, which started with two stations in 2009 (*Tóth et al.,* 2012), or various types of wide field cameras. The latter is the case of the Spanish Meteor Network dating back to 1997 (*Trigo-Rodríguez et al.,* 2001) and Polish Fireball Network operated since 2004 (*Olech et al.,* 2006). The video-based networks provide lower resolution and thus lower precision of data than photographic networks. Video cameras are usually more sensitive and capture fainter meteors, while bright bolides are saturated and hardly measurable. The quality and reliability of the data critically depends on the quality of the astrometric procedures. Amateur astronomers in many countries now operate video cameras optimized for fainter meteors (e.g. *Molau and Rendtel,* 2009; *SonotaCo,* 2009). The International Meteor Organization (www.imo.net) plays an important role in coordinating and popularizing these activities. Occasionally, faint-meteor cameras also capture meteorite falls, as was the case of Slovenian and Croatian Meteor Networks (*Spurný et al.,* 2010, *Šegon et al.,* 2011) and the professional CAMS system in California (*Jenniskens et al.,* 2014).

Optical cameras can be used as meteor spectrographs by putting a diffraction element (prism or grating) in front of the lens. Such objective spectrographs do not need any slit, since meteors are line objects and their monochromatic images form the spectrum. Large format photographic film cameras with rather long focal length have been used to obtain high resolution spectra of bright bolides (Fig. 2). The description and analysis of such detailed spectra, containing over a hundred of emission lines, was published in a number of papers (e.g. *Halliday,* 1961; *Ceplecha,* 1971; *Borovička,* 1993, 1994b). High resolution photographic spectrographs are still in use at Ondřejov Observatory. A more sensitive CCD spectrograph, having, however, smaller field of view, was used during the Leonid campaign (*Jenniskens,* 2007). The Spanish Meteor Network uses video cameras to obtain low resolution spectra of bolides (*Madiedo et al.,* 2013a,b, 2014a).

Detailed light curves of bolides can be studied by radiometers, i.e. photoelectric or semiconductor devices, which do not image the bolide, but measure the total scattered sky light as a function of time (Fig. 3). Radiometers are part of AFOs and DAFOs in Central Europe and of AFOs in Australia and provide the light curves with either 500 or 5000 samples per second (*Spurný and Ceplecha,* 2008). Radiometric light curves of superbolides are also detected globally as a byproduct of satellite-based systems constructed for different purposes. These systems provide global detection of superbolides, with radiometric measurements establishing bolide time and radiant power (*Tagliaferri et al.*, 1994). Total energy of the event may also be estimated under a number of assumptions. A complimentary system provides estimates of the location and in some cases the velocity/altitude or fragmentation height of superbolides. Meteorological and scientific satellites have also detected bolides in flight or their remnant dust/aerosol clouds (*Borovička and Charvát*, 2009, *Klekociuk et al., 2005, Rieger et al., 2014*).

In addition to systems exploiting the electromagnetic emission of bolides, the atmospheric shock waves produced by fireballs may be detected (*Edwards et al.*, 2011). At large distances from the bolide, these shocks are detectable as infrasound, which is sound below ~20 Hz. Bolide infrasound is detectable by microbarographs, which are instruments able to record coherent pressure amplitudes as low as one part in $10^8$ of the ambient atmospheric pressure. When three or more such sensors are deployed within a region of order 1 km in size, the resulting sensor array can efficiently distinguish coherent sources from noise based on cross-correlation of the pressure signals and measured signal arrival direction and elevation (*Christie and Campus*, 2010). Since ~2000 the International Monitoring System (IMS) of the Comprehensive-Test Ban Treaty Organization (CTBTO) has operated infrasound arrays on a global scale. As of 2014 some 47 stations (of a projected network of 60) are operating and provide nearly global detection capability for kiloton-scale superbolides.

In contrast to infrasound observations which directly detect the very small pressure perturbation from bolides at ranges of thousands of kilometers, seismographs respond to the bolide air waves coupled to the solid Earth. The low efficiency of seismo-acoustic coupling implies that airwaves are normally only directly detectable seismically within a few hundred kilometers of a bolide. For very energetic bolides, surface coupled waves may also be produced (*Edwards et al.*, 2008). The principle advantage of seismic bolide measurements are the tens of thousands of seismic stations operated globally which provide dense coverage in some areas sometimes allowing multiple seismo-acoustic detections of a single bolide which can provide information on trajectory and fragmentation points (e.g. *Borovička and Kalenda*, 2003; *Pujol et al.*, 2006).

Radar may also be used in several modes to characterize bolides. Direct detection of the radar head echo associated with the fireball, whereby radio waves are reflected off electrons in the region of the fireball head can be used to estimate the velocity of the bolide as well as compute its trajectory (*Brown et al.*, 2011) and in principle its mass. The main limitation of this observational mode is the large power-aperture (or large radar cross section) needed to detect such head echoes which effectively limit detectability to small regions in the atmosphere near the radar (*Kero et al.*, 2011). The ionization trail left behind from a bolide is more easily detected than the corresponding head echo (*Ceplecha et al.*, 1998) - this provides some limited information on the location of the bolide and may be used to place constraints on its ablated mass.

Recently, weather radars (*Fries and Fries,* 2010) have been shown to be effective at detecting the debris plume from a bolide in the form of material in dark flight. These Doppler radars are able to detect macroscopic-sized fragments as well as finer dust drifting to Earth, typically at altitudes <20 km (and in some instances at altitudes as low as just a few kilometers) after a bolide. They provide a means to confirm deposition of meteorites on the ground together with fall location as they probe the terminal stages of dark flight.

## 3.2 Data analysis

Once analyzed, optical data from fireball networks provide fireball trajectories, velocities and light curves. The first very important step is astrometric reduction of wide-field or all-sky images (*Borovička et al.,* 1995). In most cases, bolide trajectories can be considered to be straight. The triangulation from two or more stations can be done either with the plane intersection method (*Ceplecha,* 1987) or the straight least square method (*Borovička,* 1990; *Gural,* 2012). The trajectories of long-duration, nearly horizontal bolides are non-negligibly curved by Earth's gravity and special procedures are needed to reconstruct the trajectories of such so-called Earth-grazing fireballs (*Ceplecha* 1979, *Borovička and Ceplecha*, 1992).

To measure bolide velocities on long-exposure photographic records, the cameras are equipped with a shutter interrupting the exposure periodically (with frequencies of the order of 10 Hz). By measuring the positions of the shutter breaks or individual video images, the fireball position (length along the trajectory or height) as a function of time is obtained. This information forms the basis of dynamic analysis, i.e. determination of bolide velocity and deceleration as a function of time and, consequently, other parameters such as ablation coefficient and *dynamic mass* of the meteoroid. The parameters can be obtained by fitting the integral solution of the drag and mass loss equations first obtained by *Pecina and Ceplecha* (1983). The method assumes a single body meteoroid, i.e. no fragmentation. The method of *Gritsevich* (2009) is an equivalent treatment. In reality, however, fragmentation is a very common and important process. A more general method allowing the determination of one fragmentation point from bolide dynamics was developed by *Ceplecha et al.* (1993). For a review, see *Ceplecha et al.* (1998).

Another method of estimating the meteoroid mass is derived from the bolide light curve. The mass obtained by the integration of the luminosity equation is called the photometric mass. The crucial parameter is the luminous efficiency, which is, unfortunately, not well known. It has been discussed in a number of papers (e.g. *Ceplecha and McCrosky,* 1976; *Pecina and Ceplecha,* 1983; *Hill et al.,* 2005; *ReVelle and Ceplecha*, 2001; *Weryk and Brown,* 2013). In general, it appears that luminous efficiency increases with meteoroid velocity and mass. It is also expected on theoretical grounds that the luminous efficiency depends on height of flight and on flow regime. These dependencies are not as important for large meteoroids because they affect mostly the beginning of the bolide. Theoretical estimates of luminous efficiency were published by Golub et al. (1996) and Nemtchinov et al. (1997).

In addition to the differential luminous efficiency found in the luminosity equation, an integral luminous efficiency is defined as the ratio of the total energy radiated by the bolide to its initial kinetic energy. The integral luminous efficiency also increases with meteoroid energy (*Brown et al.,* 2002a). To give a rough value, in a typical meteorite fall (velocity ~ 15 km/s, initial mass ~ 100 kg), both integral and differential luminous efficiency of large fragments are about 5%.

A major issue which is not explainable by uncertainties in the luminous efficiency is the large difference found between the dynamic and photometric masses of meteoroids. The underlying physical reason for these order of magnitude disparities is meteoroid fragmentation, which was not taken into account properly in early treatments. To obtain self-consistent meteoroid mass and other properties of the meteoroid, modeling its atmospheric flight including fragmentation using both dynamic and photometric data (and, if possible, also other data such as direct observation of fragments) is desirable. One such attempt was done by *Ceplecha and ReVelle* (2005). A somewhat different approach was used by *Borovička et al.* (2013a) when analyzing the Košice meteorite fall.

Besides optical data, the energy of the fireball can be estimated also from acoustic data. The information available for acoustic energy estimation is the pressure amplitude and dominant period of the pressure waveform from infrasound measurements. Due to the complex, changing nature of the atmosphere, it is not generally possible to theoretically determine initial source energy from infrasound measurements alone. Instead, empirical estimates are made using calibrations from explosive sources with known energy. As pressure amplitude is most modified in atmospheric propagation, the signal period tends to be most robust in estimating source energy. Using data derived from simultaneous infrasound and satellite measurements of bolides, *Ens et al.* (2012) found that bolide source energies E (in tons of TNT equivalent energy) could be related to average infrasound periods (in seconds) via $E=3.16\tau^{3.75}$.

These bolide-specific energy relations are nearly identical to those derived from nuclear weapons tests. These period energy estimates become less certain at small energies (small periods) and more reliable as more stations are used to provide average period estimates. Amplitude-based yields are similarly calibrated from ground explosions and bolides, but are less robust, show wide scatter and are highly sensitive to stratospheric winds. Amplitude based yields become increasingly unreliable at large ranges or large source energies (above ~ 1 kt TNT). Hence at short ranges and small energies, amplitude-based energy estimates are preferred, while for large ranges and large energies periods become more reliable.

The bolide observations can also be used to estimate the terminal mass, i.e. the mass of fragments, which survived the flight and landed as meteorites (if any). The mass of an individual well observed fragment can be computed from the atmospheric density, $\rho$, velocity, $v$, and deceleration, $dv/dt$, at the end of the luminous trajectory using the drag equation:

$$m_E = (\Gamma A)^3 \rho^3 v^6 \delta^{-2} (dv/dt)^{-3},$$

where the drag coefficient, $\Gamma$, shape factor, $A$, and meteorite density, $\delta$, must be assumed (for a sphere, $A=1.21$; $\Gamma \sim 0.6$ at the relevant heights and velocities). The landing point can be estimated by the numerical integration of the object's motion during dark flight (*Ceplecha*, 1987) without knowing its mass – just the direction of flight, and last observed height, velocity, and deceleration are needed as initial values. The largest uncertainty in this landing point estimate arises from the error in the deceleration (which is not observed directly but must be computed from the observed length as a function of time over some part of the trajectory) and also in the uncertainty in the high altitude winds, which can shift the meteorite by many kilometers (e.g. *Spurný et al.*, 2012a). Note, however, that in some cases small, not directly observable fragments, can form the bulk of the fall mass (e.g. *Borovička et al.*, 2013a, *Spurný et al.*, 2014). Their presence may be revealed when modelling the light curve. Large numbers of meteorites makes the recovery more likely, even if they are small (<< 1 kg).

Meteorites usually represent only a small part of the initial mass of the meteoriod. In some cases quite a large fraction of the total mass is deposited in the atmosphere in the form of micron-sized dust particles which avoid complete vaporization (*Klekociuk et al.,* 2005; *Borovička and Charvát,* 2009). This dust slowly sediments out of the atmosphere over timescales of months and can spread over large geographical areas (*Gorkavyi et al.,* 2013). In addition, part of the vaporized material may recondense, forming nm-sized meteoric smoke (*Borovička and Charvát,* 2009), which is distributed globally (*Megner et al.,* 2008).

Early data from fireball networks showed that there are enormous differences in fireball end heights even if the initial conditions (velocity, mass, slope of the trajectory) are similar. This is most readily interpreted as differences in physical strength among meteoroids within the fireball population. On this basis, *Ceplecha and McCrosky* (1976) created the PE criterion and classified bolides into four types based on physical strength: namely types I, II, IIIA, and IIIB. A typical meteorite dropping bolide, which a priori is expected to be near the high end of all meteoroid physical strengths is of type I and was associated by *Ceplecha and McCrosky* (1976) to stony meteoroids of densities > 3000 kg/m$^3$. Meteoroid densities are not directly measurable from bolide data; nevertheless, statistical arguments led *Ceplecha and McCrosky* (1976) to the conclusion that type II bolides correspond to carbonaceous chondrites of densities ~ 2000 kg/m$^3$ and IIIA and IIIB are two types of cometary material of densities of about 750 kg/m$^3$ and 300 kg/m$^3$, respectively (*Ceplecha,* 1988). The latter two types disintegrate high in the atmosphere and do not provide meteorites.

The bolide end height is easily measurable and the PE criterion (which derives from the symbol $\rho_E$ representing the atmospheric mass density at the end point of the fireball) is still used today. A complementary method of classification uses the apparent ablation coefficient of the fireball. It is obtained by fitting the bolide dynamics without considering fragmentation. Typical values are 0.014, 0.042, 0.1, and 0.21 s$^2$/km$^2$ (or kg/MJ) for fireball types I, II, IIIA, and IIIB, respectively (*Ceplecha,* 1988; *Ceplecha et al.,* 1998). However, as shown by *Ceplecha and ReVelle* (2005), if fragmentation is taken into account, the obtained *intrinsic ablation coefficient* is nearly the same for all four types and is quite low, about 0.005 s$^2$/km$^2$. This suggests that the composition of the material is similar in all cases (mostly silicates) and the main differences are in the bulk density, porosity and mechanical strength, which determines the degree of meteoroid fragmentation. Note that iron meteoroids are not considered in this scheme, since they are quite rare and we do not have enough bolide data to characterize them as a population. In fact, iron meteoroids are not easily recognizable without spectral observation or recovered meteorites. They have larger density than stony meteoroids (~7800 kg/m$^3$) and probably also have a larger intrinsic ablation coefficient because of lower melting temperature and higher thermal conductivity (*ReVelle and Ceplecha,* 1994).

Finally, if discrete fragmentation points are identified on the fireball trajectory, the dynamic pressure acting on the meteoroid at that point can be used to classify the meteoroid. The fragile IIIB bodies often disrupt catastrophically at pressures of several to several tens of kilopascals accompanied by conspicuous flares (*Borovička and Spurný,* 1996; *Borovička et al., 2007; Madiedo et al.,* 2014a). The type I bodies fragment usually into macroscopic pieces, often several times consequtively, at pressures from a few tenths to a few MPa (*Ceplecha et al.,* 1993; *Popova et al.,* 2011).

The heliocentric orbit of the meteoroid can be computed from the time of entry, direction of flight (radiant coordinates), and velocity. The corrections to Earth gravity and Earth rotation must be applied. The corrected radiant and velocity are called *geocentric*. An analytical method of orbit computation was presented by *Ceplecha* (1987). *Clark & Wiegert* (2011) presented a computationally more demanding numerical method and showed that Ceplecha's formulation remains valid except in rare particular cases.

## 4. METEORITE ANALYSES RELEVANT TO FIREBALL STUDIES

Meteorites, when recovered, are subject to detailed mineralogical, chemical, physical, and isotopic analyses in the laboratory. Some of them are directly relevant to fireball studies because they provide either independent estimates of some parameters or supplementary information.

In interplanetary space, meteoroids are bombarded by energetic particles of solar and galactic cosmic rays. The interaction of cosmic rays with atoms in the meteoroid leads to the formation of atomic nuclei which are otherwise only rare or absent in the meteoric material, namely noble gases and short-lived radionuclides (*Verchovsky and Sephton,* 2005; *Eugster et al.,* 2006). Cosmic rays penetrate a few meters inside the meteoroid. The measurement of the concentration of selected nuclides, in particular $^{60}$Co, $^{10}$Be, $^{21}$Ne, and $^{22}$Ne, can be used to estimate the pre-atmospheric radius of the meteoroid and the depth of the measured sample inside the meteoroid. Note, however, that the calculations are model-dependent and rely on derived nuclide production rates (*Leya and Masarik,* 2009). The correspondence with the fireball-determined pre-atmospheric mass of the meteoroid is not always good (*Popova et al.*, 2011).

The cosmogenic nuclides can be also used to estimate the *cosmic-ray exposure* (CRE) age of the meteoroid. This is the time for which the meteoroid was exposed to cosmic rays, i.e. the time elapsed since the meteoroid was excavated from a deeper depth in its parent body. Lunar meteorites and carbonaceous chondrites have the shortest CRE ages (about 1 Myr on average) while the CRE ages of iron meteorites are the longest (hundreds of Myr), see *Eugster et al.,* (2006).

The measurement of physical properties of meteorites is also important for modeling the meteoroid flight in the atmosphere. Densities and porosities of meteorites of various types were published in a series of papers (*Macke et al.,* 2011 and references therein). Thermal conductivities (*Opeil et al.,* 2012) and heat capacities (*Consolmagno et al.,* 2013) were also measured. Other measurements of these quantities were obtained by *Beech et al.* (2009) and their application to meteor physics was discussed. The measurements of tensile and compressive strengths of meteorites are still rare because a destructive analysis is needed. Published values were compiled by *Popova et al.* (2011) and *Kimberley and Ramesh* (2011).

## 5. INSTRUMENTALLY OBSERVED METEORITE FALLS

Instrumental observations of fireballs which produce meteorite falls and finds are of great scientific interest and importance because meteorites provide us with a surviving physical record of the formation of our Solar System, and a direct link to their parent bodies. But most meteorites are also unique - as geological materials - in that they come with virtually no spatial context to aid us in interpreting that record. Reliable orbital information for meteorite falls is known for only 22 cases. This is a tiny fraction of the tens of thousands of meteorites which are known. For this reason, every new fireball producing meteorite with precise orbital data gives us invaluable information.

Similarly important is the study of processes accompanying the atmospheric flight of the meteoroid producing a meteorite fall. From every new instrumentally documented fall we learn very much to help refine our methods and models. On the other hand the known properties of the meteorite, such as density, mass, shape, composition, structure, etc., facilitate reverse calibration of other fireball data which do not produce meteorites. This provides information on the physical properties of meteoroids not likely contained in our meteorite collections.

Basic data about all instrumentally documented meteorite falls (22 cases so far) are shown in Tables 1–3, where details about their orbital data, atmospheric trajectory data and meteorite data are collected. When comparing the data, it must be realized that the data were obtained under various circumstances (day/night, different ranges to the bolide) and by widely different techniques. The techniques are listed in the last column of Table 1. Only in a minority of cases (8) was the fireball trajectory fully determined from professional photographic or video networks aimed at fireball observations. In 7 cases, only casual videos and photographs were available. In 4 cases, a combination of the two or video cameras aimed at fainter meteors were used. Tagish Lake and Almahata Sitta fireballs were not imaged in flight from the ground at all. Satellite data and dust cloud images were used instead. In the case of Almahatta Sitta, the orbit was precisely known from pre-impact observation of the meteoroid. To some extent also the atmospheric trajectory could be determined from that data.

It is difficult to link specific meteorites with individual parent asteroids with some certainty. The orbit measured for any particular meteorite producing fireball is heavily evolved from the original orbit of the parent asteroid. As a result, linking individual fireballs with specific asteroids is generally not possible. Rather, classes of meteorites and source regions in the main asteroid belt can be statistically associated using the orbit distribution of many meteorites and models of main belt – near Earth asteroid delivery (*Bottke et al.,* 2002a).

We tried to evaluate the quality of various aspects of fireball data in the 13$^{th}$ column of Table 1. The trajectory (direction of flight, geographical coordinates, heights), dynamics (deceleration along trajectory), photometry (absolute brightness, shape of the light curve) and heliocentric orbit (dependent on the precision of the radiant, initial velocity, and time of the fireball) were evaluated on scale 1 – 5 (1- means between 1 and 2 etc.). Overall, Tagish Lake, Buzzard Coulee, and Sutter's Mill have less reliable data.

The number of digits for orbital elements in Table 1 is given so that the published error is on the last digit. Most orbits are of Apollo type, only one is Aten type. The inclinations are mostly low (the median value is about 5 degrees) but go up to 32 degrees. All aphelia lie within the orbit of Jupiter, although the Tisserand parameters ($T_J$) of Maribo and Sutter's Mill are lower than 3. These two fireballs had the largest entry velocity ($V_\infty$ in Table 2). The other values given in Table 2 are: the observed terminal velocity ($V_E$), the best estimate of the initial mass of the meteoroid ($m_\infty$), the maximal absolute magnitude of the fireball ($M_{max}$), the slope of the trajectory (the slope is changing along the trajectory due to Earth's curvature, so only rough values are given), the observed beginning height ($H_B$), the height of maximum brightness ($H_{max}$), the terminal height ($H_E$), total length of the trajectory ($L$), duration of the fireball ($T$), total energy ($E$) in kilotons of TNT (1 kt TNT = $4.185 \times 10^{12}$ J) and the maximal encountered dynamic pressure ($P_{max}$). The observed beginnings of the fireballs (and thus also the lengths and durations) strongly depend on the technique and in the case of casual records also on pure chance. We therefore list in some cases also values from visual observations. The terminal heights are affected by observation effects to a lesser extent because the drop of brightness at the end is usually steeper.

It can be seen that meteorites were observed to fall from meteoroids of a wide range of masses, causing fireballs different by orders of magnitude in terms of energy and brightness. At the lower end, there were meteoroids of initial masses of only a few dozens of kg causing fireballs of absolute magnitude of about −10 or slightly more. Some meteorite falls were produced by large (>meter-sized) meteoroids associated with super-bolide events which occur globally every two weeks (*Brown et al.,* 2002a, 2013a). As a result, only a very small number of these events have detailed atmospheric flight observed. Our sample includes several cases with good dynamic and photometric data (such as Benešov, Košice, Chelyabinsk). In these instances we can obtain insight into the internal structure of the pre-atmospheric meteoroid, for comparison with the physical structure of asteroids as determined from other kind of observations. One of the measurable values is the mechanical strength expressed by the dynamic pressure at fragmentation. The enlarged sample confirms the conclusions of *Popova et al.* (2011) about the low strengths of interplanetary meteoroids and small asteroids. The relatively large value for Chelyabinsk (18 MPa) concerns only a very minor part of the body. The majority of the material was, in fact, destroyed under 1– 5 MPa (*Borovička et al.,* 2013b).

Data on types and masses of recovered meteorites are compiled in Table 3. Some of the large meteoroids disrupted heavily in the atmosphere and produced large numbers of small meteorites. This was the case for not only all three carbonaceous chondrite meteorites (which were made of relatively weak rock) and two mineralogically heterogeneous bodies (Almahata Sitta and Benešov) but also of ordinary chondrite bodies like Košice. The opposite example is Carancas discussed in the next section. But there are bodies also in our sample, which did not fragment very heavily (e.g. Morávka, Příbram and, in particular, Neuschwanstein). Chelyabinsk fragmented extensively; nevertheless, one large (600 kg) piece survived intact to the ground.

In the next section we discuss some notable meteorite falls caused by large meteoroids in more detail.

TABLE 1. Instrumentally observed meteorite falls – orbital data

| Name | Geocentric radiant | | $V_g$ (km/s) | a (AU) | e | q (AU) | Q (AU) | ω (deg) | Ω (deg) | i (deg) | $T_J$ | Quality* T/D/P/O | Techniques† |
|---|---|---|---|---|---|---|---|---|---|---|---|---|---|
| | $α_g$ (deg) | $δ_g$ (deg) | | | | | | | | | | | |
| **Příbram** | 192.338 | 17.467 | 17.431 | 2.401 | 0.6711 | 0.78951 | 4.012 | 241.750 | 17.7915 | 10.482 | 3.16 | 1, 2-, 3, 1 | P |
| **Lost City** | 315.0 | 39.1 | 9.2 | 1.66 | 0.417 | 0.967 | 2.35 | 161.0 | 283.8 | 12.0 | 4.14 | 1 ,1 ,1-, 2 | P |
| **Innisfree** | 6.66 | 66.21 | 9.4 | 1.872 | 0.4732 | 0.986 | 2.758 | 177.97 | 317.52 | 12.27 | 3.81 | 1, 1-, 1, 1 | P |
| **Benešov** | 227.617 | 39.909 | 18.081 | 2.483 | 0.6274 | 0.92515 | 4.040 | 218.370 | 47.001 | 23.981 | 3.08 | 1, 1, 1-, 1 | P |
| **Peekskill** | 209.0 | -29.3 | 10.1 | 1.49 | 0.41 | 0.886 | 2.10 | 308.0 | 17.030 | 4.9 | 4.47 | 2-, 3, 3-, 2- | CV, CP |
| **Tagish Lake** | 90.4 | 29.6 | 11.3 | 1.98 | 0.55 | 0.884 | 3.08 | 224.4 | 297.901 | 2.0 | 3.66 | 3-, 4-, 4, 3- | CTP, I, S, Sa |
| **Morávka** | 250.1 | 54.96 | 19.6 | 1.85 | 0.47 | 0.9823 | 2.71 | 203.5 | 46.258 | 32.2 | 3.70 | 2, 2, 3-, 2- | CV, I, S, Sa |
| **Neuschwanstein** | 192.33 | 19.54 | 17.51 | 2.40 | 0.670 | 0.7929 | 4.01 | 241.20 | 16.827 | 11.41 | 3.16 | 1, 1-, 2, 1 | P, S, I, PE |
| **Park Forest** | 171.8 | 11.2 | 16.1 | 2.53 | 0.680 | 0.811 | 4.26 | 237.5 | 6.1156 | 3.2 | 3.08 | 2-, 4, 3, 3 | CV, I, S, Sa |
| **Villalbeto de la Peña** | 311.4 | -18.0 | 12.9 | 2.3 | 0.63 | 0.860 | 3.7 | 132.3 | 283.671 | 0.0 | 3.30 | 2, 2-, 3-, 2- | CV ,CP, I, S |
| **Bunburra Rockhole** | 80.73 | -14.21 | 6.743 | 0.8529 | 0.2427 | 0.6459 | 1.05991 | 210.04 | 297.595 | 8.95 | 6.88 | 1, 1, 1, 1 | P, PE |
| **Almahata Sitta** | 348.1 | 7.6 | 6.45 | 1.3082 | 0.31206 | 0.89996 | 1.7164 | 234.449 | 194.101 | 2.542 | 4.93 | 3, 4-, 4-, 1‡ | I, Sa, T, CTP |
| **Buzzard Coulee** | 290.1 | 77.0 | 14.2 | 1.25 | 0.228 | 0.9612 | 1.53 | 211.3 | 238.937 | 25.0 | 5.04 | 3-, 4, 5, 3- | CV ,I |
| **Maribo** | 124.7 | 19.7 | 25.8 | 2.48 | 0.807 | 0.479 | 4.5 | 279.2 | 297.122 | 0.11 | 2.91 | 2-, 3, 2, 3 | P, CV ,R ,PE |
| **Jesenice** | 159.9 | 58.7 | 8.3 | 1.75 | 0.431 | 0.9965 | 2.51 | 190.5 | 19.196 | 9.6 | 4.01 | 2-, 3, 2-, 2- | P, V, PE, I, S |
| **Grimsby** | 242.6 | 54.97 | 17.9 | 2.04 | 0.518 | 0.9817 | 3.09 | 159.9 | 182.956 | 28.1 | 3.50 | 1, 2, 1-, 1 | V, I, R, RD |
| **Košice** | 114.3 | 29.0 | 10.3 | 2.71 | 0.647 | 0.957 | 4.5 | 204.2 | 340.072 | 2.0 | 3.02 | 2-, 2-, 2, 3 | CV ,PE, I, S |
| **Mason Gully** | 148.36 | 9.00 | 9.322 | 2.556 | 0.6158 | 0.98199 | 4.130 | 19.00 | 203.214 | 0.895 | 3.14 | 1, 1, 1, 1 | P, PE |
| **Križevci** | 131.22 | 19.53 | 14.46 | 1.544 | 0.521 | 0.7397 | 2.35 | 254.4 | 315.55 | 0.640 | 4.30 | 1-, 2, 2, 1- | P, PE, V |
| **Sutter's Mill** | 24.0 | 12.7 | 26.0 | 2.59 | 0.824 | 0.456 | 4.7 | 77.8 | 32.77 | 2.4 | 2.81 | 3, 4, 4-, 3- | CV,CP,I,S,RD |
| **Novato** | 268.1 | -48.9 | 8.21 | 2.09 | 0.526 | 0.9880 | 3.2 | 347.37 | 24.9414 | 5.5 | 3.56 | 2-, 2-, 3, 2 | V, I, CP |
| **Chelyabinsk** | 333.82 | 0.28 | 15.14 | 1.72 | 0.571 | 0.738 | 2.70 | 107.67 | 326.459 | 4.98 | 3.97 | 1-, 2, 2, 1- | CV, CP,I,S,Sa |

All angular orbital values are in J2000.0 equinox

* Quality coefficient describes the achieved reliability in determination of the atmospheric trajectory (T), dynamics (D), light curve and photometry (P) and heliocentric orbit (O) – 1-best, 5-worst

† Instrumental techniques used for the data acquisition: P – Dedicated photographic network, CV – Casual video, V – Dedicated video network, I – Infrasound, S – Seismic, Sa – Satellite, R – Meteor radar, PE – Fast photoelectric photometer, T – Telescope (pre-atmospheric observation of the meteoroid), RD – Weather Doppler radar, CP – Casual photograph, CTP – Casual trail photograph

‡ the most precise orbit thanks to the pre-atmospheric observations, however very limited precision in the atmospheric trajectory, dynamics and photometry determination

TABLE 2. Instrumentally observed meteorite falls – bolide data

| Name | Date (UT) (yyyy/mm/dd) | Time* (UT) | $V_\infty$ (km/s) | $V_E$ (km/s) | $m_\infty$ (kg) | $M_{max}$ | Slope† (deg) | $H_B$ (km) | $H_{max}$ (km) | $H_E$ (km) | L (km) | T (s) | E (kT) | $P_{max}$ (MPa) |
|---|---|---|---|---|---|---|---|---|---|---|---|---|---|---|
| **Příbram** | 1959/04/07 | 19:30:20 | 20.886 | - | 1 300 | -19 | 43 | 97.8 | 46. | 22/13§ | 111. | 5. | 0.07 | >0.9 |
| **Lost City** | 1970/01/04 | 2:14 | 14.150 | 3.4 | 163 | -11.6 | 38 | 85.9 | 31.7 | 19.5 | 107.9 | 8.9 | 0.004 | 2.8 |
| **Innisfree** | 1977/02/06 | 2:17:38 | 14.54 | 2.7 | 51 | -12.1 | 68 | 62.4 | 36. | 19.9 | 45.9 | 4.09 | 0.001 | 3.7 |
| **Benešov** | 1991/05/07 | 23:03:44 | 21.256 | 5.0 | 3 500 | -19.5 | 81 | 91.675 | 24.40 | 19.227 | 73.41 | 3.9 | 0.2 | 8.4 |
| **Peekskill** | 1992/10/09 | 23:48 | 14.72 | 5 | 10 000 | -16 | 3 | 46/80§ | - | 34/30§ | 253/800§ | 20/45§ | 0.26 | 1.0 |
| **Tagish Lake** | 2000/01/18 | 16:43:42 | 15.8 | 9 | 56 000 | -22 | 16 | - | 34.4 | 29 | - | - | 1.7 | 2.2 |
| **Morávka** | 2000/05/06 | 11:51:51 | 22.5 | 3.8 | 1 500 | -20 | 20 | 45.7/80§ | - | 21.2 | 69.3/168§ | 3.4 | 0.09 | 5 |
| **Neuschwanstein** | 2002/04/06 | 20:20:13.5 | 20.95 | 2.4 | 300 | -17.2 | 49 | 84.95 | 21. | 16.04 | 90.6 | 5.3 | 0.016 | 10.9 |
| **Park Forest** | 2003/03/27 | 5:50:26 | 19.5 | - | 11 000 | -21.7 | 61 | 82. | 29 | <18. | 73 | 5. | 0.5 | 7.4 |
| **Villalbeto de la Peña** | 2004/01/04 | 16:46:45 | 16.9 | 7.8 | 600 | -18 | 29 | 47/85§ | 27.9 | 22.2 | 50/130§ | 9§ | 0.02 | 5.1 |
| **Bunburra Rockhole** | 2007/07/20 | 19:13:53.24 | 13.365 | 5.68 | 22 | -9.6 | 31 | 62.83 | 36.3 | 29.59 | 64.65 | 5.68 | 0.0005 | 0.9 |
| **Almahata Sitta** | 2008/10/07 | 2:45:40 | 12.4 | - | 40 000 | < -19.7 | 19 | - | 37 | 32.7 | - | - | 0.73 | 0.9 |
| **Buzzard Coulee** | 2008/11/21 | 0:26:40 | 18.0 | - | 8 000 | - | 66 | 81.3 | - | 17.6 | 69.6 | 4 | 0.31 | - |
| **Maribo** | 2009/01/17 | 19:08:32.73‡ | 28.3 | - | 1 500 | -19 | 31 | 112.4 | 37.1 | 30.5 | 158.0 | 6.5 | 0.14 | 3 |
| **Jesenice** | 2009/04/09 | 0:59:40.5 | 13.78 | - | 170 | -15. | 59 | 88. | 26.4 | 18. | 82. | 6.6 | 0.004 | 3.9 |
| **Grimsby** | 2009/09/26 | 1:02:58.40 | 20.91 | 3.1 | 30 | -14.8 | 55 | 100.5 | 39. | 19.6 | 94. | 6.04 | 0.002 | 3.6 |
| **Košice** | 2010/02/28 | 22:24:46.6 | 15.0 | 4.5 | 3 500 | -18. | 60 | 68.3 | 36. | 17.4 | 59. | 4.5 | 0.1 | 6 |
| **Mason Gully** | 2010/04/13 | 10:36:12.68‡ | 14.648 | 4.1 | 14 | -10.5 | 54 | 83.46 | 35.78 | 23.84 | 73.76 | 5.97 | 0.0004 | 1.5 |
| **Križevci** | 2011/02/04 | 23:20:39.9 | 18.21 | 4.5 | 50 | -13.7 | 65 | 98.10 | 31. | 21.81 | 84 | 5.3 | 0.002 | 3.6 |
| **Sutter's Mill** | 2012/04/22 | 14:51:12 | 28.6 | 19 | 40 000 | -18.3 | 26 | 90.2 | 47.6 | 30.1 | 135 | 5 | 4 | - |
| **Novato** | 2012/10/18 | 2:44:29.9 | 13.67 | - | 80 | -13.8 | 19 | 94.4 | 36. | 22¶ | ~220 | ~18 | 0.002 | - |
| **Chelyabinsk** | 2013/02/15 | 3:20:21 | 19.03 | 3.2 | 12x10⁶ | -28. | 18 | 95.1 | 30. | 12.6 | 272. | 17. | 500 | 18 |

* Time of the beginning of the fireball unless otherwise noted
† Angle between the trajectory and horizontal plane
‡ Time of the fireball is given for the maximum brightness of the fireball which was reached in the short flare near the end of luminous trajectory
§ Value from visual observations
¶ Extrapolated value from non-calibrated image not used for the trajectory determination

TABLE 3. Instrumentally observed meteorite falls – meteorite data

| Name | Country[*] | Day/Night fall[†] | No of meteorites | Meteorite type(s) | Recovered mass (kg) | Estimated terminal mass (kg) | Meteorite coordinates[‡] (deg) | | First meteorite recovery[§] | References[¶] |
|---|---|---|---|---|---|---|---|---|---|---|
| | | | | | | | Longitude | Latitude | | |
| **Příbram** | CZE | N(E) | 4 (5[**]) | H5 | 5.6 (~7[**]) | 80. | 14.242 E | 49.658 N | A | 1, 13, 15 |
| **Lost City** | USA | N(E) | 4 | H5 | 17.342 | 25. | 95.092 W | 36.005 N | DS | 2, 8, 19 |
| **Innisfree** | CAN | N(E) | 9 | L5 | 4.58 | 4.9. | 111.337 W | 53.415 N | DS | 3, 4, 19 |
| **Benešov** | CZE | N | 4 | LL3.5,H5,PA | 0.0116 | 200-300[††] | 14.632 E | 49.766 N | DS | 46, 7, 10, 5 |
| **Peekskill** | USA | N(E) | 1 | H6 | 12.4 | - | 73.917 W | 41.283 N | I | 6, 9 |
| **Tagish Lake** | CAN | T(M) | ~500 | C2 | ~10 | 1300 | 134.227 W | 59.727 N | I | 12, 11, 20 |
| **Morávka** | CZE | D | 6 | H5-6 | 1.40 | 100 | 18.538 E | 49.592 N | I | 14, 15 |
| **Neuschwanstein** | DEU/AUT | N(E) | 3 | EL6 | 6.226 | 20 | 10.813 E | 47.525 N | DS | 13, 16 |
| **Park Forest** | USA | N | >100 | L5 | ~30 | - | 87.68 W | 41.48 N | I | 17 |
| **Villalbeto de la Peña** | ESP | D(E) | 36 | L6 | 5.2 | 13 | 4.66 W | 42.81 N | I | 21, 18, 39 |
| **Bunburra Rockhole** | AUS | N(M) | 3 | Euc-anom | 0.339 | 1.1 | 129.19 E | 31.35 S | DS | 35, 22 |
| **Almahata Sitta (2008 TC3)** | SDN | N(M) | >650 | Ure-Anom, EL,EH, H, L, LL, CB, R | ~11 | 39 | 32.40 E | 22.75 N | DS | 23, 28, 24, 29, 44 |
| **Buzzard Coulee** | CAN | N(E) | ≥2500 | H4 | >50 | - | 109.85 W | 53.00 N | TS | 25, 27 |
| **Maribo** | DNK | N(E) | 1 | CM2 | 0.026 | ~10 | 11.467 E | 54.762 N | TS | 38, 26, 34 |
| **Jesenice** | SVN | N | 3 | L6 | 3.61 | 10-30 | 14.050 E | 46.421 N | A | 30, 31 |
| **Grimsby** | CAN | N(E) | 13 | H4-6 | 0.215 | ~5 | 79.617 W | 43.200 N | A | 32 |
| **Košice** | SVK | N | 218 | H5 | 11.28 | 500[††] | 21.16 E | 48.76 N | DS | 40 |
| **Mason Gully** | AUS | N(E) | 1 | H5 | 0.0245 | 1. | 128.215 E | 30.275 N | DS | 36, 47 |
| **Križevci** | HRV | N | 1 | H6 | 0.291 | 0.3[‡‡] | 16.590 E | 46.039 N | DS | 48, 33 |
| **Sutter's Mill** | USA | D | 77 | CM2 | 0.943 | - | 120.93 W | 38.81 N | RD | 37 |
| **Novato** | USA | N(E) | 6 | L6 | 0.363 | - | 122.57 W | 38.11 N | A | 45 |
| **Chelyabinsk** | RUS | D(M) | >1000 | LL5 | >730[‡‡] | ~10000[§§] | 61.00 E | 54.86 N | I | 41, 42, 43 |

[*] ISO 3166-1 alpha-3 code  
[†] D – Day, N – Night (dark sky), T – Twilight, E – Evening, M – Morning  
[‡] If more meteorites than one were recovered, approximate middle of the strewn field is given

§ Circumstances of the recovery of the first meteorite: DS – find from dedicated search based on detailed computation of trajectory and dark flight from instrumental records; TS – find from trial search based on rough analysis of visual and instrumental records; I – casual find independent on bolide observation; A – casual find after an alert was issued on possible meteorite fall in the given area; RD – recovery based on analysis of the Doppler radar data

¶ References to values in all three tables

** Including one lost meteorite

†† Vast majority of terminal mass was in small gram sized meteorites

‡‡ Only one larger piece expected; in addition, gram-sized meteorites could be deposited in a large area

§§ One very large piece (~ 600 kg), otherwise mostly small meteorites

**References:** [1] Ceplecha (1961); [2] McCrosky et al. (1971); [3] Halliday et al. (1978); [4] Halliday et al. (1981); [5] Spurný (1994); [6] Brown et al. (1994); [7] Borovička and Spurný (1996); [8] Ceplecha (1996); [9] Ceplecha et al. (1996); [10] Borovička et al. (1998); [11] Brown et al. (2000); [12] Brown et al. (2002b); [13] Spurný et al. (2003); [14] Borovička et al. (2003); [15] Borovička and Kalenda (2003); [16] ReVelle et al. (2004); [17] Brown et al. (2004); [18] Llorca et al. (2005); [19] Ceplecha and ReVelle (2005); [20] Hildebrand et al. (2006); [21] Trigo-Rodríguez et al. (2006); [22] Bland et al. (2009); [23] Jenniskens et al. (2009); [24] Borovička and Charvát (2009); [25] Hildebrand et al. (2009); [26] Keuer et al. (2009); [27] Milley (2010); [28] Shaddad et al. (2010); [29] Bischoff et al. (2010); [30] Spurný et al. (2010); [31] Bischoff et al. (2011); [32] Brown et al. (2011); [33] Šegon et al. (2011); [34] Haack et al. (2012); [35] Spurný et al. (2012a); [36] Spurný et al. (2012b); [37] Jenniskens et al. (2012); [38] Spurný et al. (2013); [39] Bischoff et al. (2013); [40] Borovička et al. (2013a); [41] Borovička et al. (2013b); [42] Brown et al. (2013a); [43] Popova et al. (2013) [44] Horstmann and Bischoff (2014); [45] Jenniskens et al. (2014); [46] Spurný et al. (2014); [47] Borovička et al., in preparation, 2014; [48] This work (to be published)

# 6. NOTABLE METEORITE PRODUCING FIREBALLS

## 6.1. Almahata Sitta (2008 TC$_3$) and Benešov : heterogeneous falls

The Almahata Sitta meteorite fall (*Jenniskens et al.,* 2009) is unique in many aspects. It was the first meteorite fall that was predicted in advance – still the only case of this kind. The corresponding meteoroid/asteroid, designed 2008 TC$_3$, was discovered 19 hours before it entered the terrestrial atmosphere. During that time interval, numerous astrometric observations were made from various observatories around the world. Thanks to these data, a precise orbit was computed and the impact trajectory was predicted over the Nubian Desert in the Sudan with a precision of better than 1 km. Moreover, photometric and spectroscopic observations of 2008 TC$_3$ were also made and helped to characterize the body before the atmospheric entry.

There was not enough time to setup any fireball cameras in the region of impact. Consequently, the available data of the actual fireball are scarce. Nevertheless, the data clearly showed that the meteoroid was a fragile body and disrupted relatively high in the atmosphere. In accordance with this, a large number (~ 600) of mostly small (0.2 – 379 g) meteorites were recovered but no big meteorite was found. Quite surprisingly, the meteorites were of various mineralogical types. This fact completely changed our paradigm that one meteorite fall produces meteorites of one type and that (undifferentiated) asteroids have a certain mineralogical composition, although polymict meteorite breccias were known before (*Bischoff et al.,* 2010).

*Shaddad et al.* (2010) described over 600 recovered meteorites. Their total mass was 10.7 kg; the individual masses ranged from 0.2 – 379 g. The locations of all meteorites were carefully documented but only a few of the meteorites were analyzed in detail. From the classified ones, the majority was ureilites, i.e. a relatively rare type of achondrite. However, enstatite chondrites and ordinary chondrites were found as well. *Kohout et al.* (2010) measured magnetic susceptibility of 62 meteorites from that sample. In 25 cases he found anomalous values suggesting that the meteorites are not ureilites. *Bischoff et al.* (2010) analyzed a different sample of 40 meteorites from undocumented locations within the Almahata Sitta strewn field and found many different types. The mineralogical measurements of 110 meteorites by various authors were summarized by *Horstmann and Bischoff* (2014). From them, 75 were ureilites or ureilite-related, 28 were enstatite chondrites (both EH and EL), 5 were ordinary chondrites (H, L, LL), one was carbonaceous chondrite (CB) and one was a previously unknown type of chondrite related to R-chondrites. Such a variety of meteorite types within one fall is unprecedented. Naturally, the question arises if all meteorites were really part of 2008 TC$_3$ and did not come from unrelated meteorite fall(s) with overlapping strewn fields. The main arguments for common origin are as follows: (1) All meteorites are similarly fresh looking. (2) There is no indication that the non-ureilitic meteorites were located in a specific part of the strewn field (*Shaddad et al.,* 2010). (3) The presence of short-lived radionuclides in two non-ureilitic meteorites is evidence for a recent fall, consistent with the association with 2008 TC$_3$ (*Bischoff et al.,* 2010). (4) The analysis of noble gases and radionuclides in two other non-ureilitic meteorites provided the same pre-atmospheric radius and the same cosmic ray exposure age (20 Myr) as for ureilitic meteorites (*Meier et al.,* 2012). So, although not definitely proven, it is very likely that most, if not all, of the various meteorite types really belonged to the same fall and that 2008 TC$_3$ was therefore a highly heterogeneous body. Although the ureilitic lithology was prevailing, completely different lithologies were present as well. It is worth noting that none of the foreign lithologies was found to be directly embedded within the ureilitic meteorites. It therefore seems that the chondritic material was only loosely bound within the asteroid.

From the foregoing it may seem that asteroid 2008 TC$_3$ was a rubble pile, i.e. a conglomerate of rocks bound together only by mutual gravity. However, other data do not support that view. *Scheirich et al.* (2010) and *Kozubal et al.* (2011) determined the shape and rotation of 2008 TC$_3$ from pre-impact photometry. The asteroid was an elongated body with axial ratio of approximately 1:0.54:0.36. It was in an excited rotation state with period of rotation of 99.2 s and period of precession of 97.0 s. The absolute dimensions are uncertain due to uncertainties in albedo. The V-band albedo of selected meteorites was measured as 0.046 ± 0.005 by *Jenniskens et al.* (2009) and as 0.11 by *Hiroi et al.* (2011). The asteroid dimensions can be, in principle, determined also from its mass and density. The mass can be inferred from bolide energy, since the entry velocity is well known. The bolide energy was estimated as $(6.7 \pm 2.1) \times 10^{12}$ J from infrasound detection in Kenya (*Jenniskens et al.*, 2009). The radiated energy measured by US Governement sensors was $4 \times 10^{11}$ J (*Jenniskens et al.*, 2009), which translates to a total energy $(2.7 - 5.1) \times 10^{12}$ J, depending on the value of luminous efficiency (*Borovička and Charvát*, 2009). The densities of most of the 45 meteorites measured by *Shaddad et al.* (2010) were around 2800 kg m$^{-3}$ but values as high as 3430 kg m$^{-3}$ and as low as 1590 kg m$^{-3}$ were found. The bulk density of the asteroid could in principle be even lower if significant macroporosity was present. Combining the possible range of all values, *Kozubal et al.* (2011) concluded that the most probable mean size was 4.1 m, mass 50,000 kg, albedo 0.05 and bulk density 1800 kg m$^{-3}$. They did not use the shape model of *Scheirich et al.* (2010). *Kohout at al.* (2011) found a mean density of five Almahata Sitta ureilities of 3100 kg m$^{-3}$. Considering the higher albedo values of *Hiroi et al.* (2010) as more reliable and assuming significant macroporosity, they concluded that the mass of 2008 TC$_3$ was between 8,000 – 27, 000 kg only. *Welten et al.* (2010) estimated the radius × density on the basis of radionuclide measurements to 3000 ± 300 kg m$^{-2}$. We consider the following parameters as the most likely: size 6.6 × 3.6 × 2.4 m, volume 22 m$^3$, mean radius 1.74 m, mass 40,000 kg, bulk density 1800 kg m$^{-3}$, porosity almost 50% (using grain density 3500 kg m$^{-3}$ from *Kohout et al.*, 2011), albedo 0.049, bolide energy 3.1× 10$^{12}$ J, integral luminous efficiency 13%. This implies that the albedo of *Hiroi et al.* (2010) and the energy derived from infrasound estimates were overestimated. The higher albedo would lead to a smaller mass and thus an even higher conflict with infrasonic energy. On the other hand, if that energy were true, the resulting asteroid density would be too high in comparison with some of the meteorites and considering the atmospheric behavior of the body or an unrealistically low albedo would be needed.

The fast rotation of 2008 TC$_3$ means that the centrifugal force at the surface exceeded self-gravity. So, 2008 TC$_3$ was not a classical rubble pile held together only by gravity. Nevertheless, as shown by *Sánchez and Scheeres* (2014), a cohesive strength of only 25 Pa would be sufficient to bind the body together. Such strength could be provided by van der Waals forces between constituents grains (*Sánchez and Scheeres*, 2014). However, such a small mechanical strength would lead to the disintegration of the body just at the beginning of atmospheric entry at heights above 100 km. The bolide observations by US Government Sensors revealed that the bolide exhibited three flares; the middle of them was the brightest and occurred at a height of 37 km (*Jenniskens et al.*, 2009). The Meteosat satellite data confirmed the maximum at 37 km and revealed two earlier flares at 45 km and 53 km (*Borovička and Charvát*, 2009). Finally, the distribution of meteorites is consistent with their release at 37 km height (*Shaddad et al.*, 2010). So, the major disruption of 2008 TC$_3$ occurred at a height of 37 km under the dynamic pressure 0.9 MPa. This pressure is within the lower range as compared to other bodies of similar sizes (see Table 2), confirming the fragile nature of 2008 TC$_3$, but it is still much higher that the expected strength of rubble piles. The actual structure of this body remains unclear. In any case, the recovered meteorites of various physical and mineralogical properties represent only a tiny fraction of the original mass, namely the fraction that was the strongest. They were probably embedded within a matrix which mostly disappeared during the atmospheric entry. The matrix could be similar to the meteorites with the lowest measured densities, i.e. the porous fine-grained ureilites (Bischoff et al., 2010).

The reflectance spectrum of 2008 TC$_3$ was taken in the wavelength range 550 – 1000 nm (*Jenniskens et al.,* 2009). The spectrum was flat and featureless and 2008 TC$_3$ was classified as F-type. *Jenniskens et al.* (2010) searched for asteroids of similar spectra. They were not able to identify an asteroid family as the source of 2008 TC$_3$. A similar search was performed by *Gayon-Markt et al.* (2012), who also discussed the origin of 2008 TC$_3$ and concluded that it is highly improbable that the heterogeneous structure was formed by low velocity impacts in the current asteroid belt. *Horstmann and Bischoff* (2014) proposed that the material was formed in the early Solar System by a four stage process: (1) Heating and partial melting on the ureilite parent body (UPB), including basaltic magmatism. (2) An impact event that resulted in the catastrophic disruption of the UPB. (3) Rapid cooling of the released mantle material. (4) Re-accretion into smaller daughter asteroids forming ureilitic "second generation" asteroids. The foreign (chondritic) fragments were more likely incorporated in to them at the fourth stage rather than by subsequent impacts. Finally, 2008 TC$_3$ separated from one of the second generation ureilitic bodies 20 Myr ago. More recently, *Goodrich et al.* (2015) argued that the proportion of foreign clasts in 2008 TC$_3$ was not larger than in other polymict ureilites and that the same selection of materials as in other polymict ureilites was present. They hypothesized that the immediate parent of 2008 TC$_3$ was also the immediate parent of all other ureilitic meteorites. In contrast to *Horstmann and Bischoff* (2014), *Goodrich et al.* (2015) considered more likely that the foreign fragments were accreted by the 2008 TC$_3$ parent body during long periods of time.

Almahata Sitta is not the only heterogeneous meteorite fall. The second confirmed case is Benešov, a fall instrumentally observed and very well documented in 1991 (*Spurný*, 1994), including a rich bolide spectrum. The search for meteorites was unsuccessful at that time, nevertheless, several papers were devoted to the bolide analysis (e.g. *Borovička and Spurný,* 1996; *Borovička et al.,* 1998). In the spring of 2011, after a complete reanalysis of all available records (*Spurný et al.,* 2014) the meteorite search was resumed with a changed strategy – not looking for big pieces but for small ones produced in large amounts from the disruption at a height of 24 km. The new strategy was successful and four weathered meteorites with masses in the predicted range were found with metal detectors exactly in the predicted area (*Spurný et al.,* 2014). Surprisingly, one meteorite was H chondrite, one was LL chondrite and one was LL chondrite with embedded achondritic clast (the fourth meteorite could not be classified because of its small size and weathering stage). The size and location of all four meteorites exactly in the predicted area for corresponding masses, the same degree of weathering and composition consistent with the bolide spectrum along with the extremely low probability of two coincidental falls in the given area, means that almost certainly all meteorites came from the Benešov bolide**.** The heterogeneous nature of the Benešov meteoroid is supported by its early separation into smaller bodies during the atmospheric flight (*Borovička et al.,* 1998).

These findings can shed new light on some old meteorite finds. The Galim meteorite fall contained both LL and EH specimens (*Rubin,* 1997). Other examples of meteorites of different types found close to each other can be found in the Meteoritical Database, e.g. Hajmah (ureilite+L), Gao-Guenie (H+CR), and Markovka (H+L). These meteorites were traditionally classified separately, e.g. as Hajmah (a), Hajmah (b), but may in fact come from the same body. Of course, chance alignment in these non-observed falls is also possible. The study of the surprisingly rich Franconia strewn field containing various meteorite types concluded that they fell at various times (*Hutson et al.,* 2013). On the other hand, there are some meteorites containing foreign clasts on a microscopic scale. The most prominent example is the polymict microbreccia Kaidun, which contains materials of many different meteorite types in mm-sized clasts (*Zolensky and Ivanov,* 2003). Other cases have been summarized by *Bischoff et al.* (2010).

## 6.2. Příbram and Neuschwanstein meteorite pair

The Příbram meteorite fall has a special status among instrumentally recorded falls. This is not only from the fact that it was the first such case in history (it fell on April 7, 1959) and that it was for the first time when the recovered meteorites were directly linked with asteroids, but also that it was for the first time when the so called dark flight (first use of this term which was invented by Z. Ceplecha) was rigorously computed for individual meteorite trails seen on the photographs (*Ceplecha,* 1961). This method was crucial for the future recovery of several subsequent events and it played fundamental role also in the recovery of the second predicted meteorite fall in Europe, Neuschwanstein. Coincidentally, this other extraordinary case still increased the original significance of the historic Příbram fall. The spectacular Neuschwanstein bolide was recorded by the all-sky cameras of the European Fireball Network over Austria and Germany exactly 43 years after the Příbram fall. Based on the analysis of EN photographic records, three meteorites of corresponding masses were found exactly in the predicted area (*Spurný et al.,* 2003). However, the uniqueness of this case is not in the successful recovery of meteorites, but in the fact that the heliocentric orbits of both Příbram and Neuschwanstein meteoroids were almost identical, with $D_{SH}$ = 0.025 (Fig. 4). Such close similarity of orbits for two independent meteorite falls with recovered meteorites is unknown among the other 20 meteorite falls having known orbits. Only in the case of Innisfree and Ridgedale (presumed fall) meteorite falls observed in the Canadian fireball network MORP (*Halliday,* 1987) has a similarly close orbital pair observed; unfortunately meteorites were found only for the Innisfree fall. Nevertheless, this case generated the idea of the existence of meteorite streams (*Halliday et al.,* 1990), a notion strongly supported by the unique orbital similarity of the Příbram-Neuschwanstein pair. On the other hand Příbram and Neuschwanstein falls differ in meteorite composition (Příbram is H5 while Neuschwanstein is EL6) and in cosmic-ray exposure ages (Příbram is 12 Myr while Neuschwanstein is 48 Myr). Therefore this case has generated wide discussion as to whether the apparent orbital connection between these two meteoroids is real (supported by *Tóth et al.,* 2011) or only coincidental (*Pauls and Gladman,* 2005). This question is not reliably solved yet; it seems from other recent instrumental observed falls such as Almahata Sitta or Benešov, the compositional difference shouldn't be a decisive argument against the Příbram and Neuschwanstein connection.

## 6.3. Carbonaceous chondrites – the weakest meteorites

Three instrumentally recorded fireballs resulted in carbonaceous chondrite meteorite falls. These three falls (Tagish Lake, Maribo and Sutter's Mill) have several common characteristics: they all were large (multi-meter) sized initial objects and they all showed flight characteristics (early fragmentation, high end heights relative to their mass and speed) indicative of a very fragile structure. While the orbit for Tagish Lake (a C2 ungrouped unusual meteorite) is solidly asteroidal (with $T_J$ = 3.7), the orbits for Maribo and Sutter's Mill (both CM2 chondrites) are on the borderline between Jupiter family comets and asteroids. Intriguingly, Maribo and Sutter's Mill have very similar orbits, suggesting both a common and relatively recent origin on the basis of short cosmic ray exposure ages (*Jenniskens et al.,* 2012). Moreover, these two meteorite falls appear to be associated with fireballs having initial speeds in excess of 28 km/s. This is substantially higher than the next fastest recovered fall (Morávka at 22.5 km/s). As surviving mass from ablation $\propto \exp(-v^2)$, even a small increase in initial speed result in large increases in ablation. This is even more remarkable as these are friable carbonaceous chondrites. The large size of the initial objects and the high altitude of the breakup allowing fragments to decelerate gradually appears to have been critical in survival of a small terminal mass in all three cases. For Tagish Lake, the estimated mass survival fraction was <2% (*Hildebrand et al.,* 2006) while for Sutter's Mill it is <0.001% (*Jenniskens et al,* 2012). While the orbits of Maribo and Sutter's Mill suggest a very recent, common origin, the CRE age of Sutter's Mill of 0.082 Ma (*Nishiizumi et al.,* 2014) places it in the 0.1-0.3 Ma CRE age grouplet of predominantly CM2 meteorites (*Caffee and Nishiizumi,* 1997) while the CRE age of Maribo at 0.8-1.4 Ma (*Haack et al.,* 2012) suggests they are not immediately related.

## 6.4. Carancas – a rare monolithic meteoroid

The Carancas meteorite fall occurred near noon local time on September 15, 2007 in the Andes Altiplano of Peru near the border with Bolivia. Remarkably, this fall produced an impact crater some 14 m in diameter, apparently the result of the hypersonic impact by a stony meteorite (*Tancredi et al.,* 2009). If the impact velocity is low, the size of the formed structure (impact pit/crater) is comparable to the impactor size (e.g. *Petaev,* 1992; *Mukhamednazarov,* 1999), whereas for high-velocity impacts the size of the crater is much larger than the impactor. The prevalence of low velocity meteorite impacts emphasizes the ubiquity of fragmentation during ablation of stony meteoroids. This is a well known feature of most meteorite producing fireballs, which show breakup at altitudes indicating that the mean strength of the parent meteoroid is of order 0.1 – 1 MPa (*Popova et al.,* 2011). No optical recordings of the fireball were obtained, but airwaves from the fireball were recorded by an infrasound station located less than 100 km from the crater. From analysis of near-field infrasound, *Le Pichon et al.* (2008) and *Brown et al.* (2008) independently derived constraints on the trajectory of the associated fireball suggesting a steep (>45 degree) entry angle and radiant azimuth located to the East. Using the constraint that the orbit is unlikely to be trans-Jovian and hence placing limits on the entry speed given the infrasonically estimated radiant, *Brown et al.* (2008) and *Borovička and Spurný* (2008) independently estimated that the associated meteoroid experienced >15 MPa ram pressure during flight and was initially most likely a few meters in size. Presuming that the crater formed because the meteoroid did not undergo significant fragmentation during flight (highly unusual for a stony meteoroid), this suggests the initial object was largely devoid of cracks and well described as a monolith. This result emphasizes the fact that meteoroid strengths and physical properties vary significantly, a conclusion well summarized as indicating there is no "average" meteoroid (e.g. *Ceplecha et al.,* 1998). The Carancas impact demonstrates that contrary to entry models, which predict stony meteoroids require in excess of 10 MT of mass to produce high-velocity impact craters on the ground (*Bland and Artemieva,* 2006), in rare cases much smaller stony objects can impact Earth's surface hypersonically. As Carancas is the only known example of such a small stony meteorite producing a high-velocity impact crater, it is unclear how common are strong (>10 MPa) monolithic meter-sized stony meteoroids among the NEA population. The energy of formation of the crater was estimated to be 2-3 tons of TNT equivalent ($10^{10}$ J) based on proximal blast effects (*Tancredi et al.,* 2009) as well as interpretation of the infrasonic signals (*Le Pichon et al.,* 2008; *Brown et al.,* 2008). The crater forming impact generated a surface wave seismically detected some 50 km from the crater. Based on the estimated crater yield and equivalent seismic magnitude of the surface wave the impact coupling was approximately 0.1%, the first direct measure of crater seismic coupling from an impact (*Tancredi et al.,* 2009).

### 6.5. Chelyabinsk – the largest well documented impact

The Chelyabinsk meteorite fall, which occurred in Russia on February 15, 2013, was an event in a fundamentally different category than any other meteorite fall in recent history. It was preceded by an extraordinarily bright superbolide, brighter than the Sun, and accompanied by damaging blast wave. The analysis of the infrasonic, seismic, and satellite data showed that the total energy was ~500 kT TNT, i.e. $2\times10^{15}$ J (*Brown et al.,* 2013a; *Popova et al.,* 2013). This energy is ~30 times larger than the energy of the Hiroshima atomic bomb. Although the explosion of an asteroid near the Tunguska River in Siberia in 1908 had much larger energy, estimated to be 5 – 20 MT TNT (*Vasilyev,* 1998; *Boslough and Crawford,* 2008), no meteorites were recovered. Moreover, only limited data exist about the Tunguska event, which occurred over a very remote region. The Chelyabinsk event, on the other hand, was casually recorded by many video cameras and represented a unique opportunity to study the entry of a body larger than 10 meters in size into the atmosphere. From the known energy, entry velocity, and density of the meteorites, the effective diameter of the Chelyabinsk was estimated to be 19 meters and the mass was 12,000 metric tons.

For such a large initial mass, the mass of material surviving as meteorites was quite small. There was only one large meteorite, which landed in Lake Chebarkul and was later recovered from the lake bottom having a mass of ~ 600 kg (*Popova et al.,* 2013). All other meteorites were smaller than 30 kg. Only a few meteorites larger than 1 kg were recovered, although the number of small meteorites was enormous. The total recovered mass is unknown but was probably not larger than 2 tonnes. The percentage of initial mass, which landed as macroscopic (>~ 1 cm) fragments was much smaller than in a typical meteorite fall. It was also smaller than theory predicted for an impacting asteroid of such size (*Bland and Artiemeva,* 2003, 2006).

The analysis of the atmospheric fragmentation revealed that severe destruction of the asteroid occurred between heights 39 – 30 km, under dynamic pressures of 1 – 5 MPa. At this early stage 95% of the mass was ablated and converted into dust and small (< 1 kg) fragments. The remaining 5% of the initial mass continued ablating in the form of meter-sized boulders which fragmented again at lower heights, under pressures of 10 – 18 MPa. The only large piece (mass ~2 T) which emerged withstood a maximum pressure of 15 MPa (*Borovička et al.,*

2013b), comparable in strength to the Carancas meteoroid, though representing less than 0.01% of the initial mass.

These fragmentation pressures are in the same range as found for smaller meteoroids based on observations of the associated fireballs (*Popova et al.* 2011). This confirms that there is no clear size-strength correlation among stony NEAs over the size range of centimeters to tens of meters. Since the actual strength and thus atmospheric behavior varies from case to case, we do not expect that Chelyabinsk will be representative of all bodies of similar size. Nevertheless, larger bodies are less decelerated, so they are subject to larger pressures when reaching denser atmospheric layers, which may lead to more destructive fragmentation.

Some asteroids are believed to be rubble piles/gravitational aggregates with a small strength of ~ 25 Pa due to van der Waals forces between constituent grains (*Sánchez and Scheeres,* 2014). A priori, we expect rubble pile meteoroids to separate into their constitutional parts at the very beginning of their atmospheric entry, under pressures of tens to hundreds of pascals. There is no evidence that such a separation occurred in the Chelyabinsk case. Nevertheless, significant loss of mass in the form of dust, probably from the surface layers, started early in flight at heights of >70 km, as demonstrated by the extent of the dust trail deposited in the atmosphere (*Borovička et al.,* 2013b; *Popova et al.,* 2013).

The Chelyabinsk case also vividly demonstrates the damage potential of small (10s of meters) NEAs. Although no significant damage was caused by the ground impact, the cylindrical blast wave originating at heights 25 – 35 km (*Brown et al.,* 2013a) caused structural damage (one collapsed roof, about 10% of windows broken, many large doors of factory halls fallen). The flying glass and other objects injured 1600 people (*Popova et al.,* 2013).

## 7. STATISTICS OF FIREBALLS AND SMALL ASTEROID IMPACTS

The number of meter-sized objects colliding with the Earth as a function of energy has been estimated from several independent techniques as shown in Fig. 5.

*Halliday et al.* (1996) used data from the MORP fireball network to construct a clear-sky survey, the only controlled flux survey of in-atmosphere fireball detections. The survey provided impact rates for meteoroids between a few tens of grams and a few tens of kilograms with a total area-time product roughly equivalent to one full day of global coverage. Notably he found that roughly 40% of all fireballs in this size range were associated with meteor showers and that the proportion of asteroidal (as opposed to cometary) meteoroids increased toward the top of the size range in the survey, rising to 70% of all meteoroids having masses of a few kilograms. A pronounced change in the slope also occurs at masses of a few kilograms probably associated with changes in meteoroid population/origin. The primary limitation of this survey is uncertainty in conversion of fireball brightness to meteoroid mass, though most masses are likely accurate to a factor of several.

Independent confirmation of these flux estimates has recently been provided by *Suggs et al.* (2014) who examined lunar impact flashes. Compared to *Halliday et al.* (1996) they find absolute flux numbers to be a factor of several lower in the kilogram range and the proportion of shower meteoroids to be >60%, though most of there data are in the tens of grams mass range, below the level where the *Halliday et al.* (1996) survey is complete. A larger difference is apparent at the smallest sizes, but the uncertain mass scale in both surveys may be the cause.

Due to the rarity of meter-sized impacts (which occur roughly once every two weeks over the entire Earth) ground-based optical systems are not efficient at recording large enough numbers of such events to estimate fluxes. *Brown et al.* (2002a) used data from space-based systems to detect impacts on a global scale of multi-meter sized meteoroids over an eight year period. Total impact energies were available for 300 events, though individual speeds were not. The resulting cumulative number of impacts per year ($N$) as a function of energy ($E$) – in units of kilotons of TNT = $4.185 \times 10^{12}$ J – was found to follow a power law of the form $N = 3.7\, E^{-0.9}$. This fit is appropriate to energies of 0.1 – 10 kT or diameters ranging from 1 – 6 m. An extension to this survey by *Brown et al.* (2013a) found similar values at these energies, but evidence for fluxes above the power-law curve at larger sizes. *Silber et al.* (2009) used acoustic records of impacts over a 14 year period to independently estimate fluxes in a similar size range. Their fluxes are systematically higher than the power law curve from *Brown et al.* (2002a) but in agreement within uncertainty with the revised values at larger sizes (>6 m) from *Brown et al.* (2013a).

At sizes above 10m diameter, telescopic population estimates are widely used to estimate flux. These estimates generally agree well with the extrapolated *Brown et al.* (2002a) power-law though are somewhat lower than the small-number statistic-limited estimates from bolide impacts at these larger sizes. The telescopic survey impact values have underlying uncertainties due to unknown population-wide collision probabilities at smaller sizes and poorly known albedo distribution of smaller NEAs. Given the widely differing sources of uncertainties across all the surveys shown in Fig. 5, the degree of agreement is good. We note that fluxes may also be derived from counting smaller lunar impact craters and such estimates (e.g. *Werner et al.*, 2002) agree well with the telescopically determined flux curve (*Harris*, 2013). We chose not to include the estimated flux from small lunar craters due to the controversy surrounding the role and importance of secondary craters at such small sizes (e.g. *McEwan and Bierhaus*, 2006).

Orbital information from telescopic data is only available in quantity for NEAs larger than ~10 m. Unlike the population in the centimeter to tens of centimeter size range, NEA "streams" appear to be non-existent (*Schunová et al.*, 2012), emphasizing both the longer collisional lifetimes of larger NEAs and a probable lack of cometary material among multi-meter-sized bodies. Of the several dozen meter-class impacts recorded in Earth's atmosphere with orbital and/or physical information about the strength of the impactor, the majority appear to be stony objects, with only a small number of probable, weak cometary bodies. Only 10% of this impacting population had Tisserand values below 3, emphasizing the likely dominance of asteroidal objects at these larger sizes. Among the 22 meteorite producing fireballs, 8 appear to have been meter-sized or larger prior to impact. None of these had clearly cometary orbits, though Maribo and Sutter's Mill (both CM2 Carbonaceous chondrites) have orbits similar to 2P/Encke.

## 8. OPEN QUESTIONS

We have shown that bolide observations provide information about physical and chemical properties of asteroidal and cometary fragments in the decimeter to decameter size range, about the processes occurring during their interaction with the atmosphere, including potentially hazardous effects, and about the size-frequency distribution of such events. The obtained pre-impact heliocentric orbits enable the study of likely source regions of meteorites. Nevertheless, there are still open questions that need to be answered by further observations and modeling. In this final section of this chapter, we discuss some of them.

## 8.1. Meteorites from comets

The question "do some meteorites come from comets" has been discussed for a long time (e.g. *Öpik*, 1968; *Padevět and Jakeš,* 1993; *Campins and Swindle,* 1998; *Lodders and Osborne*, 1999; *Gounelle et al.,* 2008). Some of the earlier studies were motivated by the apparent difficulty of transferring meteoroids from the asteroid belt to the Earth, a problem, which has now been solved (by orbital resonances and Yarkovsky effect, see *Morbidelli et al.,* 2002 and *Bottke et al.,* 2002b, in Asteroids III). Nevertheless, it was proposed that cometary nuclei may contain – in addition to ice and dust – also macroscopic boulders similar to carbonaceous asteroidal material, e.g. in the "icy-glue" model of *Gombosi and Houpis* (1986). Though other cometary models seem to be more probable (*Weissman and Lowry,* 2008), the presence of chondrule-like material in the samples of comet 81P/Wild 2 returned by the Stardust mission (*Nakamura et al.,* 2008) suggests that material, which formed meteorites, is present in comets as well, at least in small samples.

None of the known meteorite orbits is clearly cometary, though the orbits of Maribo and possibly also Sutter's Mill are close to the transition between cometary and asteroidal orbits. The carbonaceous chondrites, in particular types CI and CM, would be the primary candidates for cometary origin. These meteorites have been hydrated while cometary dust is anhydrous, but *Gounelle et al.* (2008) argued that hydration can occur in cometary interiors. *Gounelle et al.* (2006) computed the orbit of the Orgueil CI1 meteorite, which fell in France in 1864, and concluded that the aphelion probably lay beyond the orbit of Jupiter. Being based on visual observations and with absence of direct velocity information, the orbit cannot be completely trustworthy. *Trigo-Rodríguez et al.* (2009) observed a deeply penetrating fireball (no meteorites were found) and concluded that the orbit was similar to that of comet C/1919 Q2 Metcalf. Their paper, however, contains a numerical error in orbit computation. When corrected, the orbit is no more similar to the orbit of comet Metcalf, although the aphelion still lies beyond Jupiter. However, the orbit is highly sensitive to the value of initial velocity, which was difficult to measure in that particular case. A small change of velocity will make the orbit completely asteroidal. So, there remains no clear, unambiguous example of a cometary meteorite fall.

## 8.2 Meteorites from meteor showers

Meteor showers are caused by meteoroids of a common origin, in most cases cometary. For many showers, the meteorite survival is hampered by high entry velocity. Until recently, 30 km/s was considered a practical upper velocity limit for the occurrence of a meteorite fall (*Ceplecha et al.,* 1998). The fact that the Maribo meteorite almost reached this limit and was made from a soft material suggests that the actual limit may lie higher. Some of the low velocity meteor showers, like the 23.5 km/s Draconids originating from comet 21P/Giacobini-Zinner, contain, however, such fragile material that survival is excluded (*Borovička et al.,* 2007). On the other hand, the Taurid meteor shower (entry speeds ~ 26–30 km/s) contains both fragile and strong bodies, some of them seem to be capable to produce meteorites (*Brown et al.,* 2013b). The principal parent body of Taurids is supposed to be comet 2P/Encke, a comet on a peculiar orbit completely inside Jupiter's orbit. Both Taurids and comet 2P/Encke may be part of a broader "Taurid complex", which contains also several other showers and possibly several Near Earth Asteroids (see *Jenniskens,* 2006). Because of this it may be difficult in individual cases to link bolides directly with comet 2P/Encke.

The Geminid shower, on the other hand, is well defined and is one of the most active annual showers. The entry speed is 36 km/s and the parent body is (3200) Phaethon, which orbits in the inner Solar System and closely approaches the Sun ($q = 0.14$ AU, $a = 1.27$ AU, $i = 22°$). Recently it hasbeen classified as an active asteroid (*Jewitt,* 2012; *Jewitt et al.,* 2013). Based on the reflectance spectroscopy, asteroid (2) Pallas was identified as the likely parent body of Phaethon (*de León et al.*, 2010). Geminid meteoroids have been known to be relatively dense and strong (e.g. *Babadzhanov,* 2002; *Brown et al.,* 2013b) but only recently has it been demonstrated that a meteorite dropping Geminid could occur, though the meteorite was not found (*Spurný and Borovička*, 2013; *Spurný et al.*, in preparation). *Madiedo et al.* (2013b) presented similar observations concerning Gemind meteorite survival but their data were less robust. The recovery of a meteorite originating from Phaethon would be, undoubtedly, a major milestone.

**8.3 Meteorite streams**

The very close similarity of orbits of Příbram and Neuschwanstein meteorites (*Spurný et al.,* 2003) suggested that they may have a common origin and be part of a meteorite stream. A similar pairing was proposed earlier for the Innisfree meteorite and the Ridgedale bolide (*Halliday,* 1987). The idea of meteorite streams was also discussed from another perspective by *Lipschutz et al.* (1997). Meteorite streams are potentially formed by asteroidal collisions. The orbit of Chelyabinsk meteorites was found to be similar to the orbit of asteroid (86039) 1999 NC43, suggesting that Chelyabinsk body could be ejected from 1999 NC43 by a collision (*Borovička et al.,* 2013b). In that case, a meteorite stream could exist in Chelyabinsk orbit. However, the typical decoherence time of meteoroid streams in the Near Earth region is only $10^4 – 10^5$ yr (*Pauls and Gladman,* 2005), while the estimated collisional lifetime of asteroids is much longer (*Bottke at al.,* 2005). Meteorite streams should be therefore rare, though some may be expected to exist (*Jones and Williams,* 2007). In the Příbram-Neuschwanstein case, the search for a related shower of fainter meteors was negative (*Koten et al.,* 2014), implying that the stream, if it exists, contains only large bodies. For Chelyabinsk, the reflectance spectra of the meteorites and 1999 NC43 do not match well, so the association seems to be unlikely (Reddy et al., submitted to Icarus). Direct evidence of a meteorite stream or an association of a meteorite with its immediate parent body is therefore still missing.

Near Earth Asteroids have been also associated with (often unconvincing) meteor showers or individual fireballs, not necessarily meteorite dropping (e.g. *Babadzhanov et al.,* 2012, and *Madiedo et al.,* 2014b, and references therein). If some of these associations are real, they may indicate that the respective NEAa are in fact extinct comets and the stream was formed by cometary activity.

**8.4 Structure of meteoroids and details of their interaction with the atmosphere**

The internal structure of meteoroids and their bulk densities are still difficult to infer from bolide observation. Data interpretation is complicated by the fact, that the values of luminous efficiencies are not reliably known and the process of meteoroid ablation and fragmentation is not understood in detail. In particular, the structure and frequency of mixed-type meteoroids like Almahata Sitta and Benešov is unknown. There are also unexplained phenomena like periodic variations and high frequency flickering on fireball light curves (*Spurný and Ceplecha,* 2008, see also Fig. 3), large lateral velocities of fragments (*Borovička and Kalenda,* 2003; *Borovička et al.,* 2013b; *Stokan and Campbell-Brown,* 2014), and jet-like features on fireball images at high altitudes (*Spurný et al.,* 2000).

**Acknowledgements.** This work was supported by grant no. P209/11/1382 from GAČR, Praemium Academiae of the Czech Academy of Sciences, the Czech institutional project RVO:67985815, the NASA co-operative agreement NNX11AB76A, the Natural Sciences and Engineering Research Council of Canada, and the Canada Research Chairs program.

**REFERENCES**

Artemieva N. A. and Shuvalov V. V. (2001) Motion of a fragmented meteoroid through the planetary atmosphere. *Journal of Geophysical Research, 106,* 3297-3310.

Babadzhanov P. B. (2002) Fragmentation and densities of meteoroids. *Astronomy and Astrophysics, 384,* 317-321.

Babadzhanov P. B., Williams I. P., Kokhirova G. I. (2012) Near-Earth object 2004CK39 and its associated meteor showers. *Monthly Notices of the Royal Astronomical Society, 420*, 2546-2550.

Baldwin B. and Sheaffer Y. (1971) Ablation and breakup of large meteoroids during atmospheric entry. *Journal of Geophysical Research, 76,* 4653-4668

Beatty, K. (2014) Small Asteroid 2014 AA Hits Earth. *Sky and Telescope,* online edition http://www.skyandtelescope.com/astronomy-news/small-asteroid-2014-aa-hitsearth/

Beech M., Coulson I. M., Nie W., McCausland P. (2009) The thermal and physical characteristics of the Gao-Guenie (H5) meteorite. *Planetary and Space Science, 57,* 764-770.

Biberman L. M., Bronin S. Y., and Brykin M. V. (1980) Moving of a blunt body through the dense atmosphere under conditions of severe aerodynamic heating and ablation. *Acta Astronautica, 7,* 53-65.

Bischoff A., Horstmann M., Pack A., Laubenstein M., Haberer S. (2010) Asteroid 2008 TC3—Almahata Sitta: A spectacular breccia containing many different ureilitic and chondritic lithologies. *Meteoritics and Planetary Science, 47,* 1638-1656.

Bischoff A., Jersek M., Grau T., Mirtic B., Ott U., Kučera J., Horstmann M., Laubenstein M., Herrmann S., Řanda Z., Weber M., Heusser G. (2011) Jesenice – A new meteorite fall from Slovenia. *Meteoritics and Planetary Science, 46*, 793-804.

Bischoff A., Dyl K. A., Horstmann M., Ziegler K., Wimmer K., Young E. D. (2013) Reclassification of Villalbeto de la Peña—Occurrence of a winonaite-related fragment in a hydrothermally metamorphosed polymict L-chondritic breccia. *Meteoritics and Planetary Science, 48*, 628-640.

Bland P. A. and Artiemeva, N. A. (2003) Efficient disruption of small asteroids by Earth's atmosphere. *Nature 424,* 288–291.

Bland, P. A. and Artemieva, N. (2006) The rate of small impacts on Earth. *Meteoritics and Planetary Science*, *41,* 607–631.

Bland P. A. and 17 colleagues (2009) An anomalous basaltic meteorite from the innermost main belt. *Science, 325,* 1525-1527.

Bland P. A., Spurný P., Bevan A. W. R., Howard K. T., Towner M. C., Benedix G. K., Greenwood R. C., Shrbený L., Franchi I. A., Deacon G., Borovička J., Ceplecha Z., Vaughan D., Hough R. M. (2012) The Australian Desert Fireball Network: a new era for planetary science. *Australian Journal of Earth Sciences*, *59, * 177-187.

Borovička J. (1990) The comparison of two methods of determining meteor trajectories from photographs. *Bulletin of the Astronomical Institutes of Czechoslovakia,* 41, 391-396.

Borovička J. (1993) A fireball spectrum analysis. *Astronomy and Astrophysics, 279,* 627-645.

Borovička J. (1994a) Two components in meteor spectra. *Planetary and Space Science, 42,* 145-150.

Borovička J. (1994b) Line identifications in a fireball spectrum. *Astronomy and Astrophysics Supplement Series, 103,* 83-96.

Borovička J. (2005) Spectral investigation of two asteroidal fireballs. *Earth Moon and Planets, 97,* 279-293.

Borovička J. (2014) The analysis of casual video records of fireballs. In *Proceedings of the International Meteor Conference, Poznań, Poland 22–25 August, 2013* (M. Gyssens, P. Roggemans, P. Żołądek, eds.), pp. 101–105. International Meteor Organization.

Borovička J. and Ceplecha Z. (1992) Earth-grazing fireball of October 13, 1990. *Astronomy and Astrophysics, 257,* 323-328.

Borovička J. and Charvát Z. (2009) Meteosat observation of the atmospheric entry of 2008 $TC_3$ over Sudan and the associated dust cloud. *Astronomy and Astrophysics, 507,* 1015-1022.

Borovička J. and Kalenda P. (2003) The Morávka meteorite fall: 4 Meteoroid dynamics and fragmentation in the atmosphere. *Meteoritics and Planetary Science, 38,* 1023-1043.

Borovička J. and Spurný P. (1996) Radiation study of two very bright terrestrial bolides and an application to the comet S-L 9 collision with Jupiter. *Icarus, 121,* 484-510.

Borovička J. and Spurný P. (2008) The Carancas meteorite impact – Encounter with a monolithic meteoroid. *Astronomy and Astrophysics, 485,* L1-L4.

Borovička J., Spurný P., Keclíková J. (1995) A new positional astrometric method for all-sky cameras. *Astronomy and Astrophysics Supplement Series, 112,* 173-178.

Borovička J., Popova O. P., Nemchinov I. V., Spurný P., and Ceplecha Z. (1998) Bolides produced by impacts of large meteoroids into the Earth's atmosphere: comparison of theory with observations I. Benešov bolide dynamics and fragmentation. *Astronomy & Astrophysics, 334,* 713–728.

Borovička J., Spurný P., Kalenda P., and Tagliaferri E. (2003) The Morávka meteorite fall: 1 . Description of the events and determination of the fireball trajectory and orbit from video records. *Meteoritics and Planetary Science, 38,* 975–987.

Borovička J., Spurný P., Koten P. (2007) Atmospheric deceleration and light curves of Draconid meteors and implications for the structure of cometary dust. *Astronomy and Astrophysics, 473,* 661-672.

Borovička J., Tóth J., Igaz A., Spurný P., Kalenda P., Haloda J., Svoreň J., Kornoš L., Silber E., Brown P., Husárik M. (2013a) The Košice meteorite fall: Atmospheric trajectory, fragmentation, and orbit. *Meteoritics and Planetary Science, 48,* 1757-1779.

Borovička J., Spurný P., Brown P., Wiegert P., Kalenda P., Clark D., Shrbený L. (2013b) The trajectory, structure and origin of the Chelyabinsk asteroidal impactor. *Nature, 503,* 235-237.

Boslough M.B.E., Crawford D.A. (2008) Low-altitude airbursts and the impact threat. *International Journal of Impact Engineering, 35,* 1441–1448.

Bottke W. F., Morbidelli A., Jedicke R., Petit J.-M., Levison H. F., Michel P., Metcalfe T. S. (2002a) Debiased orbital and absolute magnitude distribution of the Near-Earth Objects. *Icarus, 156,* 399-433.

Bottke W. F., Jr., Vokrouhlický D., Rubincam D. P., Brož M. (2002b) The effect of Yarkovsky thermal forces on the dynamical evolution of asteroids and meteoroids. In *Asteroids III* (W. F. Bottke Jr., A. Cellino, P. Paolicchi, and R. P. Binzel, eds.), pp. 395-408.  The University of Arizona Press.

Bottke W. F., Durda D. D., Nesvorný D., Jedicke R., Morbidelli A., Vokrouhlický D., Levison H. F. (2005) Linking the collisional history of the main asteroid belt to its dynamical excitation and depletion. *Icarus, 179,* 63-94. Erratum: Icarus, 183, 235-236.

Bronshten, V. A. (1983) *Physics of meteoric phenomena.* D. Reidel Publ. Comp., Dordrecht, Holland. 356 pp. Original: Nauka, Moscow, 1981

Brown P. G., Ceplecha Z., Hawkes R. L., Wetherill G. W., Beech M., and Mossman K. (1994) The orbit and atmospheric trajectory of the Peekskill meteorite from video records. *Nature, 367,* 624–626.

Brown P. G. and 21 colleagues (2000) The fall, recovery, orbit, and composition of the Tagish Lake meteorite: A new type of carbonaceous chondrite. *Science 290,* 320-325.

Brown P., Spalding R. E., ReVelle D. O., Tagliaferri E., Worden S. P. (2002a) The flux of small near-Earth objects colliding with the Earth. *Nature, 420,* 294-296.

Brown P. G., Revelle D. O., Tagliaferri E., and Hildebrand A. R. (2002b) An entry model for the Tagish Lake fireball using seismic, satellite and infrasound records. *Meteoritics and Planetary Science, 37,* 661–676.

Brown P. G., Pack D., Edwards W. N., Revelle D. O., Yoo B. B., Spalding R. E., and Tagliaferri E. (2004) The orbit , atmospheric dynamics , and initial mass of the Park Forest meteorite. *Meteoritics and Planetary Science, 39,* 1781–1796.

Brown P., ReVelle D. O., Silber E. A., Edwards W. N., Arrowsmith S., Jackson L. E., Tancredi G., Eaton D. (2008) Analysis of a crater-forming meteorite impact in Peru. Journal of Geophysical Research (Planets), 113, id. E09007, doi:10.1029/2008JE003105.

Brown P., Weryk R. J., Kohut S., Edwards W. N., Krzeminski Z. (2010) Development of an all-sky video meteor network in Southern Ontario, Canada : The ASGARD System. *WGN, Journal of the International Meteor Organization, 38,* 25-30.

Brown P., McCausland P. J. A., Fries M., Silber E., Edwards W. N., Wong D. K., Weryk R. J., Fries J., and Krzeminski Z. (2011) The fall of the Grimsby meteorite–I: Fireball dynamics and orbit from radar, video, and infrasound records.  *Meteoritics and Planetary Science, 46*, 339-363.

Brown P. G. and 32 colleagues (2013a) A 500-kiloton airburst over Chelyabinsk and an enhanced hazard from small impactors. *Nature 503*, 238-241.

Brown P., Marchenko V., Moser D. E., Weryk R., Cooke W. (2013b) Meteorites from meteor showers: A case study of the Taurids. *Meteoritics and Planetary Science, 48,* 270-288.

Caffee M., and Nishiizumi K. (1997) Exposure Ages of Carbonaceous Chondrites-II. *Meteoritics and Planetary Science*, *32*, A26.

Campins H., Swindle T. D. (1998) Expected characteristics of cometary meteorites. *Meteoritics and Planetary Science, 33,* 1201-1211.

Ceplecha Z. (1961) Multiple fall of Příbram meteorites photographed. 1. Double-station photographs of the fireball and their relations to the found meteorites. *Bulletin of the Astronomical Institutes of Czechoslovakia, 12,* 21-47.

Ceplecha Z. (1971) Spectral data on terminal flare and wake of double-station meteor No. 38421 (Ondřejov, April 21, 1963). *Bulletin of the Astronomical Institutes of Czechoslovakia, 22,* 219-304.

Ceplecha Z. (1979) Earth-grazing fireballs. Tthe daylight fireball of Aug. 10, 1972. *Bulletin of the Astronomical Institutes of Czechoslovakia*, *30,* 349-356

Ceplecha Z. (1986) Photographic fireball networks. In *Asteroids, Comets, Meteors II* (C.-I. Lagerkvist, H. Rickman, B. A. Lindblad, H. Lundstedt, eds.) , pp. 575-582. Astronomiska Observatoriet Uppsala, Sweden.

Ceplecha Z. (1987) Geometric, dynamic, orbital and photometric data on meteoroids from photographic fireball networks. *Bulletin of the Astronomical Institutes of Czechoslovakia, 38,* 222-234.

Ceplecha Z. (1988) Earth's influx of different populations of sporadic meteoroids from photographic and television data. *Bulletin of the Astronomical Institutes of Czechoslovakia, 39,* 221-236.

Ceplecha Z. (1996) Luminous efficiency based on photographic observations of the Lost City fireball and implications for the influx of interplanetary bodies onto Earth. *Astronomy and Astrophysics, 311,* 329–332.

Ceplecha Z. (2007) Fragmentation model analysis of the observed atmospheric trajectory of the Tagish Lake fireball. *Meteoritics and Planetary Science, 42,* 185–189.

Ceplecha Z. and McCrosky R. E. (1976) Fireball end heights - A diagnostic for the structure of meteoric material. *Journal of Geophysical Research, 81,* 6257-6275.

Ceplecha Z. and Rajchl J. (1965) Programme of fireball photography in Czechoslovakia. *Bulletin of the Astronomical Institutes of Czechoslovakia, 16,* 15-22.

Ceplecha Z. and ReVelle D. O. (2005) Fragmentation model of meteoroid motion, mass loss, and radiation in the atmosphere. *Meteoritics and Planetary Science, 40,* 35-50.

Ceplecha Z., Ježková M., Boček J., Kirsten T., Kiko J. (1973) Data on three significant fireballs photographed within the European Network in 1971. *Bulletin of the Astronomical Institutes of Czechoslovakia, 24,* 13-21

Ceplecha Z., Spurný P., Borovička J., Keclíková J. (1993) Atmospheric fragmentation of meteoriods. *Astronomy and Astrophysics, 279,* 615-626.

Ceplecha Z., Brown P. G., Hawkes R. L., Wetherill G. W., Beech M., and Mossman K. (1996) Video observations, atmospheric path, orbit and fragmentation record of the fall of the Peekskill meteorite. *Earth, Moon and Planets, 72,* 395–404.

Ceplecha Z., Borovička J., Elford W. G., Revelle D. O., Hawkes R. L., Porubčan V., and Šimek M. (1998) Meteor Phenomena and Bodies. *Space Science Reviews, 84,* 327-471.

Ceplecha Z., Spalding R. E., Jacobs C. F., ReVelle D. O., Tagliaferri E., and Brown P. G. (1999). Superbolides. In *Meteoroids 1998* (W. J. Baggaley and V. Porubčan, eds.), pp. 37–54. Astronomical Institute of the Slovak Academy of Sciences.

Christie D.R. and Campus P. (2010) The IMS Infrasound Network: Design and Establishment of Infrasound stations. In *Infrasound Monitoring for Atmospheric Studies*, (A. Le Pichon, E. Blanc and A. Hauchecorne eds.), pp. 29-77.

Clark D. L., Wiegert P. A. (2011) A numerical comparison with the Ceplecha analytical meteoroid orbit determination method. *Meteoritics and Planetary Science,* 46, 1217-1225.

Close S., Brown P., Campbell-Brown M., Oppenheim M., and Colestock P. (2007) Meteor head echo radar data: Mass-velocity selection effects. *Icarus, 186,* 547-556.

Consolmagno G. J., Schaefer M. W., Schaefer B. E., Britt D. T., Macke R. J., Nolan M. C., Howell E. S. (2013) The measurement of meteorite heat capacity at low temperatures using liquid nitrogen vaporization. *Planetary and Space Science, 87,* 146-156.

Cooke W. J., Moser D. E. (2012) The status of the NASA All Sky Fireball Network. In *Proceedings of the International Meteor Conference, Sibiu, Romania, 15-18 Sep, 2011* (M. Gyssens, P. Roggemans, eds.), pp. 9-12.

Daubar I. J., McEwen A. S., Byrne S., Kennedy M. R., Ivanov B. (2013) The current martian cratering rate. *Icarus, 225,* 506-516.

de León J., Campins H., Tsiganis K., Morbidelli A., Licandro J. (2010) Origin of the near-Earth asteroid Phaethon and the Geminids meteor shower. *Astronomy and Astrophysics, 513,* id. A26 (7pp.)

Dodd, R. T. (1981) *Meteorites, a petrologic-chemical synthesis*. Cambridge Univ. Press. 368 pp.

Edwards W. N., Eaton D. W. and Brown P. G. (2008) Seismic observations of meteors: Coupling theory and observations. *Reviews of Geophysics, 46*, id. 4007, 1-21

Edwards W.N. (2010) Meteor Generated Infrasound: Theory and Observation. In *Infrasound Monitoring for Atmospheric Studies* (A. Le Pichon, E. Blanc and A. Hauchecorne eds.), pp. 361-414.

Ens T. A., Brown P. G., Edwards W. N., and Silber E. A. (2012) Infrasound production by bolides: A global statistical study. *Journal of Atmospheric and Solar-Terrestrial Physics, 80,* 208-229.

Eugster O., Herzog G. F., Marti K., Caffee M. W. (2006) Irradiation records, cosmic-ray exposure ages, and transfer times of meteorites. In *Meteorites and the Early Solar System II*. (D. S. Lauretta and H. Y., McSween Jr., eds.) University of Arizona Press. pp. 829–851.

Fairén A. G. and 15 colleagues (2011) Meteorites at Meridiani Planum provide evidence for significant amounts of surface and near-surface water on early Mars. *Meteoritics and Planetary Science, 46,* 1832-1841.

Fessenkow V. G., Huzarski R. G., and La Paz L. (1954) On the origin of meteorites. *Meteoritics, 1,* 208-227.

Fries M., Fries J. (2010) Doppler weather radar as a meteorite recovery tool. *Meteoritics and Planetary Science, 45,* 1476-1487.

Gayon-Markt J., Delbo M., Morbidelli A., Marchi S. (2012) On the origin of the Almahata Sitta meteorite and 2008 TC$_3$ asteroid. *Monthly Notices of the Royal Astronomical Society, 424,* 508-518.

Genge M. J. and Grady M. M. (1999) The fusion crusts of stony meteorites: Implications for the atmospheric reprocessing of extraterrestrial. materials. *Meteoritics and Planetary Science, 34,* 341-356.

Golub' A. P., Kosarev I. B., Nemchinov I. V., and Shuvalov V. V. (1996) Emission and ablation of a large meteoroid in the course of its motion through the Earth's atmosphere. *Solar System Research, 30,* 183-197.

Gombosi T. I., Houpis H. L. F. (1986) An icy-glue model of cometary nuclei. *Nature, 324,* 43-44

Goodrich C. A., Hartmann W. K., O'Brien D. P., Weidenschilling S. J., Wilson L., Michel P., and Jutzi M. (2015) Origin and history of ureilitic material in the solar system: The view from asteroid 2008 TC$_3$ and the Almahata Sitta meteorite. *Meteoritics and Planetary Science,* in press, doi: 10.1111/maps.12401

Gorkavyi N., Rault D. F., Newman P. A., Silva A. M., Dudorov A. E. (2013) New stratospheric dust belt due to the Chelyabinsk bolide. *Geophysical Research Letters, 40*, 4728-4733.

Gounelle M., Spurný P., Bland P. A. (2006) The orbit and atmospheric trajectory of the Orgueil meteorite from historical records. *Meteoritics and Planetary Science, 41,* 135-150.

Gounelle M., Morbidelli A., Bland P. A., Spurný P., Young E. D., Sephton M. (2008) Meteorites from the Outer Solar System? In *The Solar System Beyond Neptune* (M. A. Barucci, H. Boehnhardt, D. P. Cruikshank, and A. Morbidelli, eds.), University of Arizona Press, Tucson, pp.525-541.

Gritsevich M. I. (2009) Determination of parameters of meteor bodies based on flight observational data. *Advances in Space Research, 44,* 323-334.

Gural P. S. (2012) A new method of meteor trajectory determination applied to multiple unsynchronized video cameras. *Meteoritics and Planetary Science, 47,* 1405-1418.

Haack H., Grau T., Bischoff A., Horstmann M., Wasson J., Sørensen A., Laubenstein M., Ott U., Palme H., Gellissen M., Greenwood R. C., Pearson V. K., Franchi I. A., Gabelica Z., Schmitt-Kopplin P. (2012) Maribo–A new CM fall from Denmark. *Meteoritics and Planetary Science, 47*, 30-50.

Halliday I. (1961) A Study of Spectral Line Identifications in Perseid Meteor Spectra., *Publ. Dominion Obs. Ottawa* 25, 3–16.

Halliday I. (1973) Photographic Fireball Networks. In *Evolutionary and Physical Properties of Meteoroids* (C. L. Hemenway, P. M. Millma, A. F. Cook, eds.) NASA Special Publication, 319, 1-8.

Halliday I. (1987) Detection of a meteorite 'stream' - Observations of a second meteorite fall from the orbit of the Innisfree chondrite. *Icarus, 69,* 550-556. Erratum: Icarus 72, 239

Halliday I., Blackwell A. T., Griffin A. A. (1978) The Innisfree meteorite and the Canadian camera network. *Journal of the Royal Astronomical Society of Canada*, 72, 15-39.

Halliday I., Griffin A., and Blackwell A. T. (1981) The Innisfree meteorite fall – A photographic analysis of fragmentation, dynamics and luminosity. *Meteoritics, 16,* 153–170.

Halliday I., Blackwell A. T., and Griffin, A. A. (1989a). The typical meteorite event, based on photographic records of 44 fireballs. *Meteoritics, 24,* 65–72.

Halliday I., Blackwell A. T., and Griffin A. A. (1989b) The flux of meteorites on the earth's surface. *Meteoritics, 24*, 173-178.

Halliday I., Blackwell A. T., Griffin A. A. (1990) Evidence for the existence of groups of meteorite-producing asteroidal fragments. *Meteoritics, 25*, 93-99.

Halliday I., Griffin A. A., Blackwell A. T. (1996) Detailed data for 259 fireballs from the Canadian camera network and inferences concerning the influx of large meteoroids. *Meteoritics and Planetary Science*, 31, 185-217.

Harris A. (2013) The value of enhanced NEO surveys, Planetary Defense Conference, IAA-PDC13-05-09

Hildebrand A. R., McCausland P. J. A., Brown P. G., Longstaffe F. J., Russell S. D. J., Tagliaferri E., Wacker J. F., Mazur M. J. (2006). The fall and recovery of the Tagish Lake meteorite. *Meteoritics and Planetary Science, 41*, 407–431.

Hildebrand A. R. and 11 colleagues (2009). Characteristics of a bright fireball and meteorite fall at Buzzard Coulee, Saskatchewan, Canada, November 20, 2008. *In Lunar and Planetary Science XL*, Abstract #2505. Lunar and Planetary Institute, Houston.

Hill K. A., Rogers L. A., Hawkes R. L. (2005) High geocentric velocity meteor ablation. *Astronomy and Astrophysics, 444,* 615-624.

Hiroi T., Jenniskens P., Bishop J. L., Shatir T. S. M., Kudoda A. M., Shaddad M. H. (2010) Bidirectional visible-NIR and biconical FT-IR reflectance spectra of Almahata Sitta meteorite samples. *Meteoritics and Planetary Science, 45,* 1836-1845.

Holsapple K. (1993) The scaling of impact processes in planetary sciences. *Annual Review of Earth and Planetary Sciences, 21,* 333-373.

Holsapple K. (2009). On the strength of the small bodies of the solar system: A review of strength theories and their implementation for analyses of impact disruption. *Planetary and Space Science 57,* 127–141.

Horstmann M. and Bischoff A. (2014) The Almahata Sitta polymict breccia and the late accretion of asteroid 2008 TC3. *Chemie der Erde / Geochemistry, 74,* 149-183.

Hueso R. and 23 colleagues. (2013). Impact flux on Jupiter: From superbolides to large-scale collisions. *Astronomy and Astrophysics, 560,* id. A55, 1–14.

Hutchison R. (2004). *Meteorites, a petrologic, chemical and isotopic synthesis*. Cambridge Univ. Press. 511 pp.

Hutson M., Ruzicka A., Timothy Jull A. J., Smaller J. E., Brown R. (2013) Stones from Mohave County, Arizona: Multiple falls in the "Franconia strewn field". *Meteoritics and Planetary Science, 48,* 365-389.

Jenniskens P. (2006) *Meteor Showers and their Parent Comets.* Cambridge Univ. Cambridge. 790 pp.

Jenniskens P. (2007) Quantitative meteor spectroscopy: Elemental abundances. *Advances in Space Research,* 39, 491-512.

Jenniskens P. and 34 colleagues (2009) The impact and recovery of asteroid 2008 $TC_3$. *Nature, 458,* 485-488

Jenniskens P., Vaubaillon J., Binzel R. P., DeMeo F. E., Nesvorný D., Bottke W. F., Fitzsimmons A., Hiroi T., Marchis F., Bishop J. L., Vernazza P., Zolensky M. E., Herrin J. S., Welten K. C., Meier M. M. M., Shaddad M. H. (2010) Almahata Sitta (=asteroid 2008 $TC_3$) and the search for the ureilite parent body. *Meteoritics and Planetary Science, 45*, 1590-1617.

Jenniskens P. and 69 colleagues, (2012) Radar-Enabled Recovery of the Sutter's Mill Meteorite, a Carbonaceous Chondrite Regolith Breccia. *Science, 338,* 1583-1587.

Jenniskens P. and 47 colleagues (2014) Fall, recovery and characterization of the Novato L6 chondrite breccia. *Meteoritics and Planetary Science 49,* 1388-1425

Jewitt D. (2012) The Active Asteroids. *Astronomical Journal, 143*, id. 66 (14pp.)

Jewitt D., Li J., Agarwal J. (2013) The Dust Tail of Asteroid (3200) Phaethon. *Astrophysical Journal Letters*, 771, id. L36 (5pp.)

Jones D. C., Williams I. P. (2008) High inclination meteorite streams can exist. *Earth Moon and Planets, 102,* 35-46.

Jones J., Brown P., Ellis K. J., Webster A. R., Campbell-Brown M., Krzemenski Z., and Weryk R. J. (2005). The Canadian Meteor Orbit Radar: System overview and preliminary results. *Planetary and Space Science 53,* 413–421.

Keay C. S. L. (1992) Electrophonic sounds from large meteor fireballs. *Meteoritics, 27,* 144-148.

Kero J., Szasz C., Nakamura T., Meisel D. D., Ueda M., Fujiwara Y., Terasawa T., Miyamoto H., Nishimura K. (2011) First results from the 2009-2010 MU radar head echo observation programme for sporadic and shower meteors: the Orionids 2009. *Monthly Notices of the Royal Astronomical Society, 416,* 2550-2559.

Keuer, D., Singer, W. and Stober G. (2009) Signatures of the ionization trail of a fireball observed in the HF, and VHF range above middle-Europe on Jan 17, 2009. In *Proceedings of the 12th workshop on Technical and Scientific Aspects of MST radar*, (Eds. N. Swarnalingham and W.K. Hocking), p. 154-158.

Kimberley J. and Ramesh K. T. (2011) The dynamic strength of an ordinary chondrite. *Meteoritics and Planetary Science, 46,* 1653-1669.

Klekociuk A. R., Brown P. G., Pack D. W., Revelle D. O., Edwards W. N., Spalding R. E., Tagliaferri E., Yoo B. B., Zagari J. (2005) Meteoritic dust from the atmospheric disintegration of a large meteoroid. *Nature, 436,* 1132-1135.

Kohout T., Jenniskens P., Shaddad M. H., Haloda J. (2010) Inhomogeneity of asteroid 2008 TC$_3$ (Almahata Sitta meteorites) revealed through magnetic susceptibility measurements. *Meteoritics and Planetary Science, 45,* 1778-1788.

Kohout T., Kiuru R., Montonen M., Scheirich P., Britt D., Macke R., Consolmagno G. (2011) Internal structure and physical properties of the Asteroid 2008 TC$_3$ inferred from a study of the Almahata Sitta meteorites. *Icarus, 212,* 697-700.

Kokhirova G. I., Borovička J. (2011) Observations of the 2009 Leonid activity by the Tajikistan fireball network. *Astronomy and Astrophysics*, 533, id. A115, 6pp.

Koten P., Vaubaillon J., Čapek D., Vojáček V., Spurný P., Štork R., Colas F. (2014) Search for faint meteors on the orbits of Příbram and Neuschwanstein meteorites. *Icarus, 239,* 244-252.

Kozubal M. J., Gasdia F. W., Dantowitz R. F., Scheirich P., Harris A. W. (2011) Photometric observations of Earth-impacting asteroid 2008 TC$_3$. Meteoritics and Planetary Science, 46, 534-542.

Lauretta D. S. and McSween Jr. H. Y., eds. (2006) *Meteorites and the Early Solar System II.* University of Arizona Press. 943 pp.

Le Pichon A. Antier K., Cansi Y., Hernandez B., Minaya E., Burgoa B., Drob D., Evers L. G., Vaubaillon J. (2008) Evidence for a meteoritic origin of the September 15, 2007, Carancas crater. *Meteoritics and Planetary Science, 43,* 1797-1809.

Leya I., Masarik J. (2009) Cosmogenic nuclides in stony meteorites revisited. *Meteoritics and Planetary Science, 44,* 1061-1086.

Lipschutz M. E., Wolf S. F., Dodd R. T. (1997) Meteoroid streams as sources for meteorite falls: a status report. *Planetary and Space Science, 45,* 517-523.

Lodders K., Osborne R. (1999) Perspectives on the comet-asteroid-meteorite link. *Space Science Reviews, 90,* 289-297.

Macke R. J., Consolmagno G. J., Britt D. T. (2011) Density, porosity, and magnetic susceptibility of carbonaceous chondrites. *Meteoritics and Planetary Science, 46,* 1842-1862.

Madiedo J. M., Trigo-Rodríguez J. M., Ortiz J. L., Castro-Tirado A. J., Pastor S., A. de los Reyes J., Cabrera-Caño J. (2013a) Spectroscopy and orbital analysis of bright bolides observed over the Iberian Peninsula from 2010 to 2012. *Monthly Notices of the Royal Astronomical Society,* 435, 2023-2032.

Madiedo J. M., Trigo-Rodríguez J. M., Castro-Tirado A. J., Ortiz J. L., Cabrera-Caño J. (2013b) The Geminid meteoroid stream as a potential meteorite dropper: a case study. *Monthly Notices of the Royal Astronomical Society, 436,* 2818-2823.

Madiedo J. M., Ortiz J. L., Trigo-Rodríguez J. M., Zamorano J., Konovalova N., Castro-Tirado A. J., Ocaña F., de Miguel A. S., Izquierdo J., Cabrera-Caño J. (2014a) Analysis of two superbolides with a cometary origin observed over the Iberian Peninsula. *Icarus, 233,* 27-35.

Madiedo J. M., Trigo-Rodríguez J. M., Ortiz J. L., Castro-Tirado A. J., Cabrera-Caño J. (2014b) Bright fireballs associated with the potentially hazardous asteroid 2007LQ19. *Monthly Notices of the Royal Astronomical Society, 443,* 1643-1650.

McCrosky R. E. and Boeschenstein H., Jr. (1965) The Prairie Meteorite Network. *SAO Special Report*, #173, 23pp.

McCrosky R. E., Posen A., Schwartz G., and Shao C.-Y. (1971) Lost City meteorites, its recovery and a comparison with other fireballs. *Journal of Geophysical Research, 76,* 4090–4108.

McEwen A. S., and Bierhaus E. B. (2006) The importance of secondary cratering to age constraints on planetary surfaces. *Annual Review of Earth and Planetary Sciences*, *34*, 535–567.

Megner L., Siskind D. E., Rapp M., Gumbel J. (2008) Global and temporal distribution of meteoric smoke: A two-dimensional simulation study. *Journal of Geophysical Research, 113,* id. D03202, 15 pp.

Meier M. M. M., Welten K. C., Caffee M. W., Friedrich J. M., Jenniskens P., Nishiizumi K., Shaddad M. H., Wieler R. (2012) A noble gas and cosmogenic radionuclide analysis of two ordinary chondrites from Almahata Sitta. *Meteoritics and Planetary Science, 47,* 1075-1086.

Milley E. P. (2010) Physical Properties of fireball-producing earth-impacting meteoroids and orbit determination through shadow calibration of the Buzzard Coulee meteorite fall., MSc. Thesis, University of Calgary, Calgary, 166 pp.

Molau S., Rendtel J. (2009) A comprehensive list of meteor showers obtained from 10 years of observations with the IMO Video Meteor Network. *WGN, Journal of the International Meteor Organization*, 37, 98-121.

Morbidelli A., Bottke W. F., Jr., Froeschlé C., Michel P. (2002) Origin and evolution of Near-Earth Objects. In *Asteroids III* (W. F. Bottke Jr., A. Cellino, P. Paolicchi, and R. P. Binzel, eds.), pp. 409-422. The University of Arizona Press.

Mukhamednazarov S. (1999) Observation of a fireball and the fall of the first large meteorite in Turkmenistan. *Astronomy Letters,* 25, 117-118.

Nakamura T., Noguchi T., Tsuchiyama A., Ushikubo T., Kita N. T., Valley J. W., Zolensky M. E., Kakazu Y., Sakamoto K., Mashio E., Uesugi K., Nakano T. (2008) Chondrulelike objects in short-period Comet 81P/Wild 2. *Science, 321,* 1664-1667.

Nemtchinov I. V., Svetsov V. V., Kosarev I. B., Golub' A. P., Popova O. P., Shuvalov V. V., Spalding R. E., Jacobs C., and Tagliaferri E. (1997) Assessment of kinetic energy of meteoroids detected by satellite-based light sensors. *Icarus, 130,* 259-274.

Nishiizumi K., Caffee M. W., Hamajima Y., Reedy R. C., Welten, K. C. (2014). Exposure history of the Sutter's Mill carbonaceous chondrite. *Meteoritics and Planetary Science*, *49*, 2056-2063

Obenberger K. S., Taylor G. B., Hartman J. M., Dowell J., Ellingson S. W., Helmboldt J. F., Henning P. A., Kavic M., Schinzel F. K., Simonetti J. H., Stovall K., and Wilson T. L. (2014) Detection of radio emission from fireballs. *Astrophysical Journal Letters, 788,* id. L26, 1-6

Oberst J., Molau S., Heinlein D., Gritzner C., Schindler M., Spurný P., Ceplecha Z., Rendtel J., Betlem H. (1998) The "European Fireball Network": Current status and future prospects. *Meteoritics and Planetary Science*, 33, 49-56.

Olech A., Żołądek P., Wiśniewski M., Krasnowski M., Kwinta M., Fajfer T., Fietkiewicz K., Dorosz D., Kowalski L., Olejnik J., Mularczyk K., Złoczewski K. (2006) Polish Fireball Network. In *Proceedings of the International Meteor Conference, Oostmalle, Belgium, 15-18 Sep, 2005* (J. Verbert, L. Bastiaens, J.-M. Wislez, C. Verbeeck, eds.), pp. 53-62.

Opeil C. P., Consolmagno G. J., Safarik D. J., Britt D. T. (2012) Stony meteorite thermal properties and their relationship with meteorite chemical and physical states. *Meteoritics and Planetary Science, 47,* 319-329.

Öpik E. J. (1950) Interstellar meteors and related problems. *Irish Astronomical Journal, 1,* 80-96.

Öpik E. J. (1968) The cometary origin of meteorites. *Irish Astronomical Journal, 8,* 185-208

Padevět V. and Jakeš P. (1993) Comets and Meteorites: Relationship (AGAIN?). *Astronomy and Astrophysics, 274,* 944-954.

Papike J. J., ed. (1998) *Planetary Materials*. Mineralogical Soc. America, Washington. 1039 pp.

Pauls A. and Gladman B. (2005) Decoherence time scales for "meteoroid streams". *Meteoritics and Planetary Science, 40,* 1241-1256.

Pecina P. and Ceplecha Z. (1983) New aspects in single-body meteor physics. *Bulletin of the Astronomical Institutes of Czechoslovakia, 34,* 102-121.

Petaev M. I. (1992) The Sterlitamak meteorite – a new crater-forming fall. *Solar System Research, 26,* 384-398.

Popova O. (2004) Meteoroid ablation models. *Earth Moon and Planets, 95,* 303-319.

Popova O. P., Strelkov A. S., and Sidneva S. N. (2007) Sputtering of fast meteoroids' surface. *Advances in Space Research, 39,* 567-573.

Popova O., Borovička J., Hartmann W. K., Spurný P., Gnos E., Nemtchinov I., and Trigo-Rodríguez J. M. (2011) Very low strengths of interplanetary meteoroids and small asteroids. *Meteoritics and Planetary Science*, 46, 1525-1550.

Popova O. P. and 59 colleagues (2013) Chelyabinsk airburst, damage assessment, meteorite recovery, and characterization. *Science 342,* 1069-1073.

Pujol J., Rydelek P., Ishihara Y. (2006) Analytical and graphical determination of the trajectory of a fireball using seismic data. *Planetary and Space Science, 54,* 78-86.

Rabinowitz D., Helin E., Lawrence K., Pravdo S. (2000) A reduced estimate of the number of kilometre-sized near-Earth asteroids. Measurement of the spatial coherence of a trapped. *Nature 403,* 165–166.

Revelle D. O. and Ceplecha Z. (1994) Analysis of identified iron meteoroids: Possible relation with M-type Earth-crossing asteroids? *Astronomy and Astrophysics,* 292, 330-336.

ReVelle D.O. and Ceplecha Z. (2001) Bolide physical theory with application to PN and EN fireballs. In: *Proceedings of the Meteoroids 2001 Conference, Kiruna, Sweden* (B.Warmbein, ed.), *ESA Special Publication vol. 495*, pp.507–512.

Revelle D. O., Brown P. G., and Spurný, P. (2004) Entry dynamics and acoustics/infrasonic/seismic analysis for the Neuschwanstein meteorite fall. *Meteoritics and Planetary Science, 39,* 1605–1626.

Rieger L. A.., Bourassa A. E., Degenstein D. A. (2014) Odin–OSIRIS detection of the Chelyabinsk meteor. *Atmospheric Measurement Techniques 7,* 777–780

Rogers L., Hill K. A., and Hawkes, R. L. (2005). Mass loss due to sputtering and thermal processes in meteoroid ablation. *Planetary and Space Science, 53,* 1341–1354.

Rubin A. E. (1997) The Galim LL/EH polymict breccia: Evidence for impact-induced exchange between reduced and oxidized meteoritic material. *Meteoritics and Planetary Science, 32,* 489-492.

Sánchez P. and Scheeres D. J. (2014) The strength of regolith and rubble pile asteroids. *Meteoritics and Planetary Science, 49,* 788-811.

Scheirich P., Ďurech J., Pravec P., Kozubal M., Dantowitz R., Kaasalainen M., Betzler A. S., Beltrame P., Muler G., Birtwhistle P., Kugel F. (2010) The shape and rotation of asteroid 2008 $TC_3$. *Meteoritics and Planetary Science, 45,* 1804-1811.

Schunová E., Granvik M., Jedicke R., Gronchi G., Wainscoat R., Abe S. (2012). Searching for the first near-Earth object family. *Icarus, 220,* 1050–1063.

Šegon D., Korlević K., Andreić Ž., Kac J., Atanackov J., Kladnik G. (2011) Meteorite-dropping bolide over north Croatia on 4th February 2011. *WGN, Journal of the International Meteor Organization,* 39, 98-99.

Shaddad M. H. and 19 colleagues (2014) The recovery of asteroid 2008 $TC_3$. *Meteoritics and Planetary Science, 45,* 1557-1589.

Silber E. A., ReVelle D. O., Brown P. G., Edwards W. N. (2009) An estimate of the terrestrial influx of large meteoroids from infrasonic measurements. *Journal of Geophysical Research, 114,* id. E08006, 8pp.

SonotaCo (2009) A meteor shower catalog based on video observations in 2007-2008. *WGN, Journal of the International Meteor Organization,* 37, 55-62.

Spurný P. (1994) Recent fireballs photographed in central Europe. *Planetary and Space Science,* 42, 157–162.

Spurný P. and Borovička J. (2013) Meteorite dropping Geminid recorded. In Meteoroids 2013, Abstract #061. Adam Mickiewicz University, Poznań, Poland. http://www.astro.amu.edu.pl/Meteoroids2013//data/abstracts.pdf

Spurný P. and Ceplecha Z. (2008) Is electric charge separation the main process for kinetic energy transformation into the meteor phenomenon? *Astronomy and Astrophysics,* 489, 449-454.

Spurný P., Betlem H., Jobse K., Koten P., van't Leven J. (2000) New type of radiation of bright Leonid meteors above 130 km. *Meteoritics and Planetary Science, 35,* 1109-1115.

Spurný P., Oberst J., Heinlein D. (2003) Photographic observations of Neuschwanstein, a second meteorite from the orbit of the Příbram chondrite. *Nature, 423,* 151-153.

Spurný P., Borovička J., Shrbený L. (2007) Automation of the Czech part of the European fireball network: equipment, methods and first results. In *Near Earth Objects, our Celestial Neighbors: Opportunity and Risk* (A. Milani, G.B. Valsecchi & D. Vokrouhlický, eds.) *IAU Symposium,* 236, 121-130.

Spurný P., Borovička J., Kac J., Kalenda P., Atanackov J., Kladnik G., Heinlein D., Grau T. (2010) Analysis of instrumental observations of the Jesenice meteorite fall on April 9, 2009. *Meteoritics and Planetary Science,* 45, 1392-1407.

Spurný P., Bland P. A., Shrbený L., Borovička J., Ceplecha Z., Singelton A., Bevan A. W. R., Vaughan D., Towner M. C., McClafferty T. P., Toumi R., Deacon G. (2012a) The Bunburra Rockhole meteorite fall in SW Australia: fireball trajectory, luminosity, dynamics, orbit, and impact position from photographic and photoelectric records. *Meteoritics and Planetary Science, 47,* 163-185.

Spurný P., Bland P. A., Borovička J. Towner M. C., Shrbený L., Bevan A. W. R., Vaughan D. (2012b). The Mason Gully meteorite fall in SW Australia: Fireball trajectory, luminosity, dynamics, orbit and impact position from photographic records. In *Asteroids, Comets, Meteors 2012, Proceedings of the conference held May 16-20, 2012 in Niigata, Japan.* LPI Contribution No. 1667, id. 6369.

Spurný, P., Borovička, J., Haack, H., Singer, W., Keuer, D., Jobse, K. (2013) Trajectory and orbit of the Maribo CM2 meteorite from optical, photoelectric and radar records, Meteoroids 2013 conference, Poznań, Poland, August 26-30, 2013, poster presentation.

Spurný P., Haloda J., Borovička J., Shrbený L., Halodová P. (2014) Reanalysis of the Benešov bolide and recovery of inhomogeneous breccia meteorites – old mystery revealed after 20 years. *Astronomy and Astrophysics, 570,* id. A39 (14pp.)

Stokan E., Campbell-Brown M. D. (2014) Transverse motion of fragmenting faint meteors observed with the Canadian Automated Meteor Observatory. *Icarus, 232,* 1-12.

Stuart J.S. (2001) A near-Earth asteroid population estimate from the LINEAR survey. *Science, 294,* 1691-1693.

Suggs R. M., Moser D. E., Cooke W. J. and Suggs R. J. (2014) The flux of kilogram-sized meteoroids from lunar impact monitoring. *Icarus, 238,* 23-36.

Svetsov V. V., Nemtchinov I. V., and Teterev A. V. (1995) Disintegration of large meteoroids in Earth's atmosphere: Theoretical models. *Icarus, 116,* 131-153.

Tagliaferri E., Spalding R., Jacobs C., Worden S.P. and Erlich A. (1994) Detection of meteoroid impacts by optical sensors in earth orbit. In *Hazards due to Comets and Asteroids* (ed. T. Gehrels), pp. 199-221 The University of Arizona Press.

Tancredi G. and 15 colleagues (2009) A meteorite crater on Earth formed on September 15, 2007: The Carancas hypervelocity impact. *Meteoritics and Planetary Science, 44,* 1967-1984.

Tóth J., Vereš P., Kornoš L. (2011) Tidal disruption of NEAs - a case of Příbram meteorite. *Monthly Notices of the Royal Astronomical Society, 415,* 1527-1533.

Tóth J., Kornoš L., Piffl R., Koukal J., Gajdoš Š., Popek M., Majchrovič I., Zima M., Világi J., Kalmančok D., Vereš P., Zigo P. (2012) Slovak Video Meteor Network - status and results: Lyrids 2009, Geminids 2010, Quadrantids 2011. In *Proceedings of the International Meteor Conference, Sibiu, Romania, 15-18 Sep, 2011* (M. Gyssens, P. Roggemans, eds.), pp. 82-84.

Trigo-Rodríguez J. M., and Llorca J. (2006) The strength of cometary meteoroids: clues to the structure and evolution of comets. *Monthly Notices of the Royal Astronomical Society, 372,* 655-660; erratum 375, 415

Trigo-Rodríguez J. M., Fabregat J., Llorca J., Castro-Tirado A., del Castillo A., de Ugarte A., López A. E., Villares F., Ruiz-Garrido J. (2001) Spanish Fireball Network: Current Status and Recent Orbit Data. *WGN, Journal of the International Meteor Organization,* 29, 139-144.

Trigo-Rodríguez J. M., Borovička J., Spurný P., Ortiz J. L., Docobo J. A., Castro-Tirado A. J., and Llorca J. (2006) The Villalbeto de la Peña meteorite fall: II. Determination of atmospheric trajectory and orbit. *Meteoritics and Planetary Science, 41,* 505–517.

Trigo-Rodríguez J. M., Madiedo J. M., Williams I. P., Castro-Tirado A. J., Llorca J., Vítek S., Jelínek M. (2009) Observations of a very bright fireball and its likely link with comet C/1919 Q2 Metcalf. *Monthly Notices of the Royal Astronomical Society, 394,* 569-576.

Vasilyev N. V. (1998) The Tunguska Meteorite problem today. *Planetary and Space Science, 46,* 129-150.

Verchovsky A. B., and Sephton M.A (2005) Noble gases in meteorites: A noble record. *Astronomy & Geophysics, 46,* 2.12-2.14.

Weissman P. R., Lowry S. C. (2008) Structure and density of cometary nuclei. *Meteoritics and Planetary Science, 43,* 1033-1047.

Welten K. C., Meier M. M. M., Caffee M. W., Nishiizumi K., Wieler R., Jenniskens P., Shaddad M. H. (2010) Cosmogenic nuclides in Almahata Sitta ureilites: Cosmic-ray exposure age, preatmospheric mass, and bulk density of asteroid 2008 $TC_3$. Meteoritics and Planetary Science, 45, 1728-1742.

Werner S. C., Harris A. W., Neukum G., Ivanov B. A. (2002) The near-Earth asteroid size-frequency distribution: A snapshot of the lunar impactor size frequency distribution. *Icarus, 156,* 287–290.

Weryk R. J., Brown P. G. (2013) Simultaneous radar and video meteors –II: Photometry and ionisation. *Planetary and Space Science, 81,* 32-47.

Zolensky M. and Ivanov A. (2003) The Kaidun microbreccia meteorite: a harvest from the inner and outer asteroid belt. *Chemie der Erde / Geochemistry, 63,* 185-246.

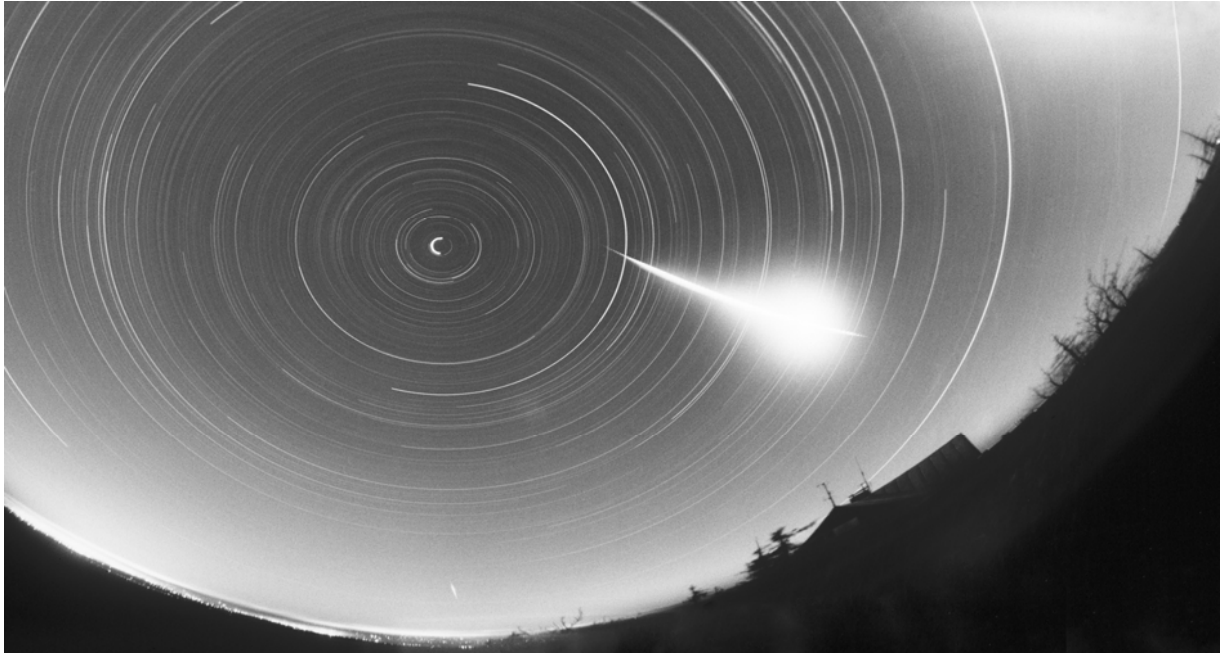

**Figure 1.** Image of a very bright bolide EN 210199 taken by the photographic all-sky camera at Lysá hora, Czech Republic. Stars form circles around the North Pole during a long exposure. Only part of the all-sky image is shown.

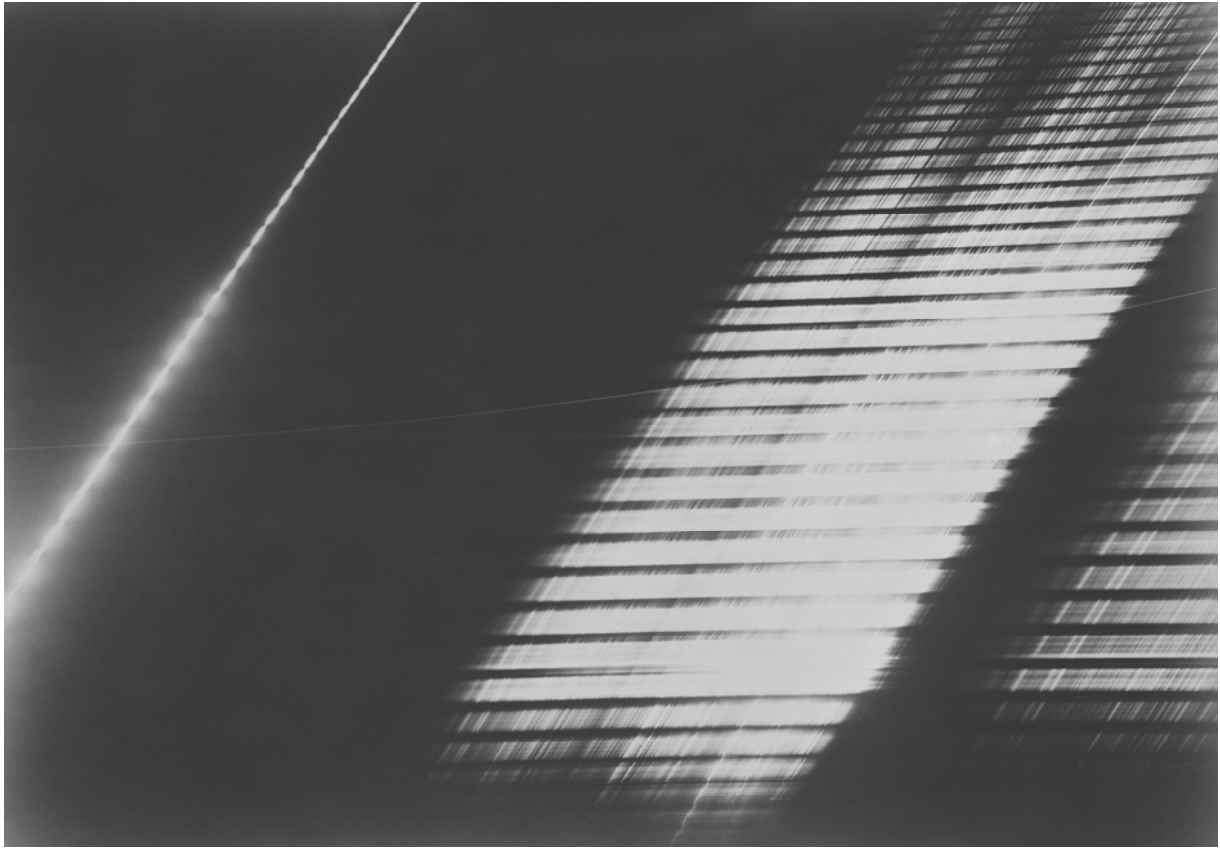

**Figure 2.** Photographic grating spectrum of a lower part of the Benešov bolide EN070591. The spectrum was taken from the Ondřejov Observatory, Czech Republic. The zero order (direct image) is on the left, the first order and part of the second order are on the right. The fireball flew from the top to the bottom. The exposure was interrupted by a rotating shutter. See Borovička and Spurný (1996) for more detailed description of the spectrum. Benešov is the only bolide with both recovered meteorites and recorded spectrum.

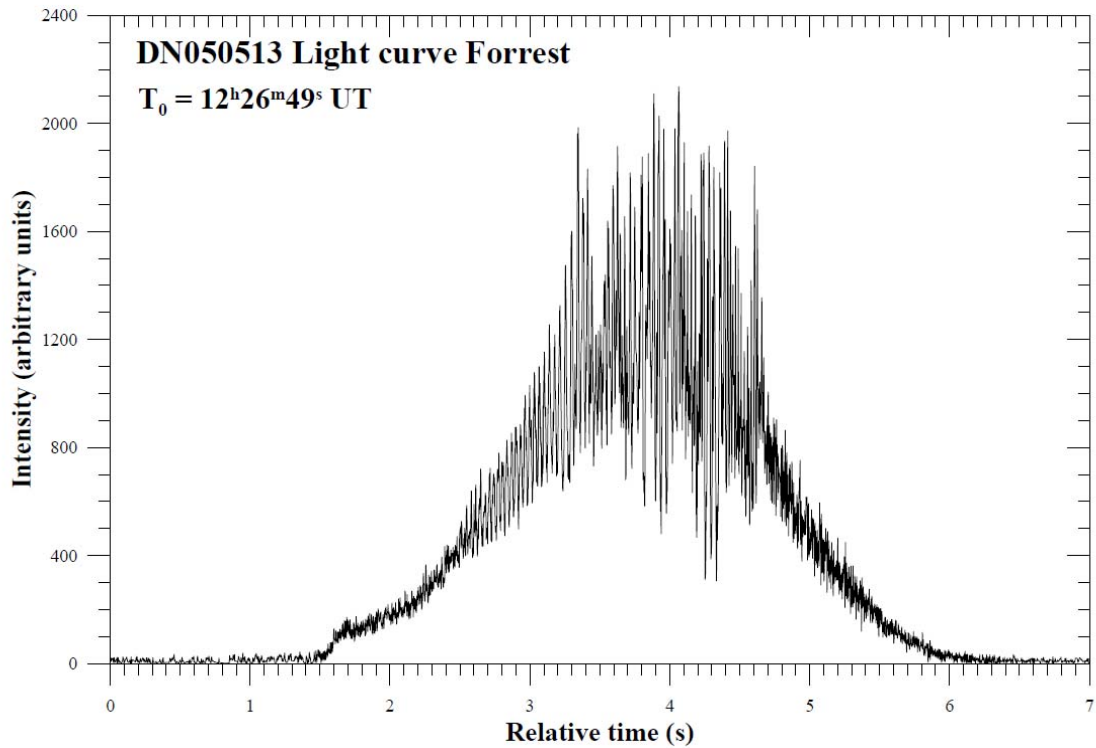

**Figure 3.** Radiometric light curve of bolide DN050513 taken by the Autonomous Fireball Observatory at station Forrest in Australia. The sampling rate was 500 Hz. The high amplitude and high frequency variations during the bright phase are real. The amplitude of the noise was much lower as it can be seen at the edges of the curve. DN050513 was a type I meteorite dropping bolide (meteorites landed in an inaccessible area and were not recovered).

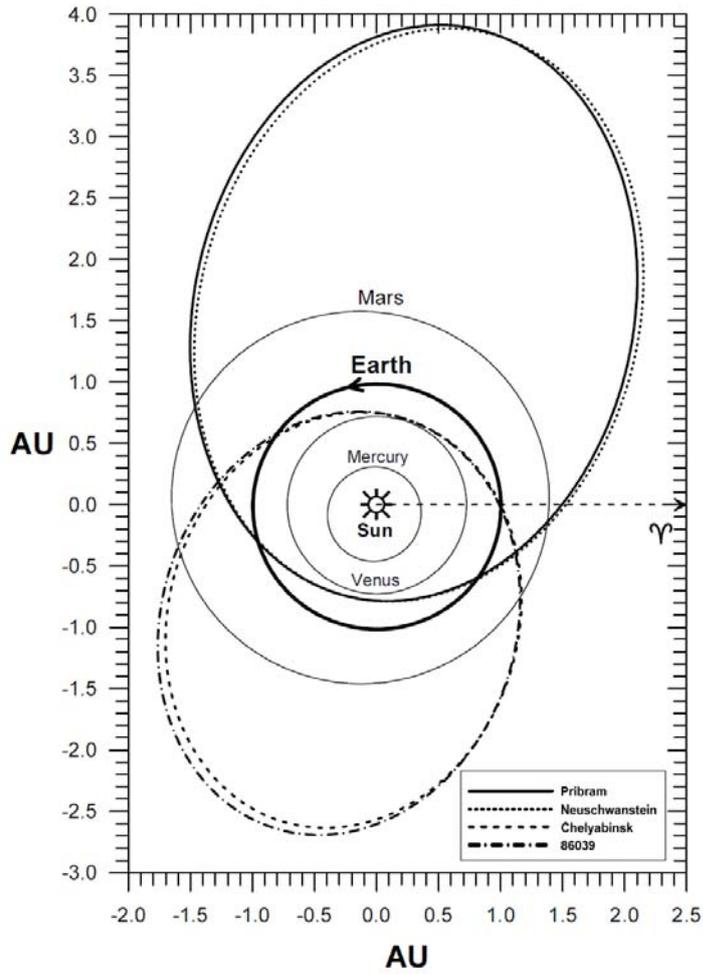

**Figure 4.** The orbits of Příbram, Neuschwanstein, and Chelyabinsk meteorites and asteroid (86039) 1999 NC43, projected into the plane of ecliptic. The arrow with sign points to vernal equinox. The orbital similarity of Příbram-Neuschwanstein and Chelyabinsk-1999 NC43 is evident.

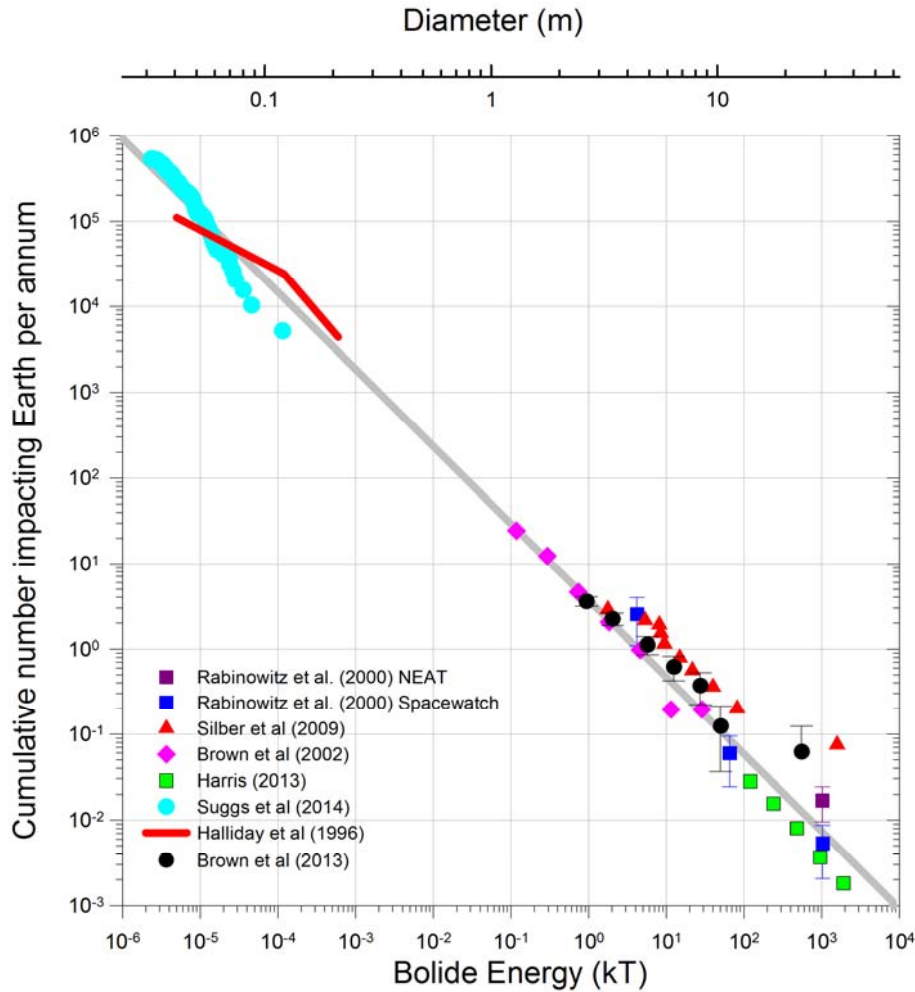

**Figure 5.** Observational estimates of the terrestrial cumulative impact flux (ordinate) as a function of impact energy (abscissa). The grey line represents the power-law fit from *Brown et al.* (2002a) based on satellite impact flash observations (pink diamonds) approximately representing the interval 0.1 – 10 kT where data is most complete. The solid circles are an update to these data as given by *Brown et al.* (2013a) with better statistics at larger sizes for energies > 1 kT.   Error bars (where shown) represent counting statistics only. These data consist of de-biased estimates of the telescopic near-Earth asteroid population and assumed average impact probabilities as given by *Rabinowitz et al.* (2000) (purple solid squares) from the NEAT survey and Spacewatch (blue squares) surveys, where diameters are determined assuming an albedo of 0.1. The LINEAR values at smaller sizes are normalized to early work which established the absolute population for diameters >100m (*Stuart,* 2001).  Also shown are the estimated impact rate from infrasonic measurements of bolide airbursts from the Air Force Technical Application Centre (AFTAC) acoustic monitoring network as reanalyzed by *Silber et al.* (2009) (red triangles). More recent telescopic de-biased estimates from data compiled from all surveys by *Harris* (2013) are shown as green squares. The cyan circles are the equivalent impact rate for the Earth as determined from lunar flashes taking into account gravitational focusing (*Suggs et al.,* 2014). Finally, the red line represents the impact rate from the photographic Meteorite Observation and Recovery Project (MORP) clear sky survey as described by *Halliday et al.* (1996).